\documentclass[fleqn,usenatbib]{mnras}

\usepackage{newtxtext,newtxmath}
\usepackage[T1]{fontenc}

\DeclareRobustCommand{\VAN}[3]{#2}
\let\VANthebibliography\thebibliography
\def\thebibliography{\DeclareRobustCommand{\VAN}[3]{##3}\VANthebibliography}

\usepackage{graphicx}	% Including figure files
\usepackage{amsmath}	% Advanced maths commands
\title[]{CAPOS: The bulge Cluster APOgee Survey IV \\Elemental Abundances of the bulge globular cluster NGC~6558}

\author[Gonz\'alez-D\'iaz et al.]{
Danilo Gonz\'alez-D\'iaz,$^{1,2}$\thanks{E-mail: danilo.gonzalez@ucn.cl}
Jos\'e G. Fern\'andez-Trincado,$^{1}$
Sandro Villanova,$^{3}$
Doug Geisler,$^{3,4,5}$
\newauthor
Beatriz Barbuy,$^{6}$
Dante Minniti,$^{7,8,9}$
Timothy C. Beers,$^{10}$
Christian Moni Bidin,$^{1}$
Francesco Mauro,$^{1}$
\newauthor
Cesar Mu\~noz,$^{4,5}$
Baitian Tang,$^{11}$
Mario Soto,$^{12}$
Antonela Monachesi,$^{4,5}$
Richard R. Lane,$^{13}$
and Heinz Frelijj$^{1}$
\\
$^{1}$Instituto de Astronom\'ia, Universidad Cat\'olica del Norte, Av. Angamos 0610, Antofagasta, Chile\\
$^{2}$Instituto de F\'isica, Universidad de Antioquia, Calle 70 52-21, Medell\'in, Colombia\\
$^{3}$Departamento de Astronom\'ia, Casilla 160-C, Universidad de Concepci\'on, Concepci\'on, Chile\\
$^{4}$Instituto Multidisciplinario de Investigaci\'on y Postgrado, Universidad de La Serena, Ra\'ul Bitr\'an 1305, La Serena, Chile  \\
$^{5}$Departamento de Astronom\'ia, Facultad de Ciencias, Universidad de La Serena. Av. Juan Cisternas 1200, La Serena, Chile \\
$^{6}$Universidade de S\~ao Paulo, IAG, Rua do Mat\~ao 1226, Cidade Universit\'aria, S\~ao Paulo 05508-900, Brazil \\
$^{7}$ Institute of Astrophysics, Facultad de Ciencias Exactas, Universidad Andr\'es Bello, Av. Fern\'andez Concha 700, Las Condes, Santiago, Chile \\
$^{8}$Vatican Observatory, V00120 Vatican City State, Italy \\
$^{9}$Departamento de Fisica, Universidade Federal de Santa Catarina, Trinidade 88040-900, Florianopolis, Brazil \\
$^{10}$Department of Physics and Astronomy and JINA Center for the Evolution of the Elements, University of Notre Dame, Notre Dame, IN 46556 USA \\
$^{11}$School of Physics and Astronomy, Sun Yat-sen University, Zhuhai 519082, China \\
$^{12}$Instituto de Astronom\'ia y Ciencias Planetarias, Universidad de Atacama, Copayapu 485, Copiap\'o, Chile\\
$^{13}$Centro de Investigaci\'on en Astronom\'ia, Universidad Bernardo O'Higgins, Avenida Viel 1497, Santiago, Chile \\
}

\date{Accepted XXX. Received YYY; in original form ZZZ}

\pubyear{2023}

\begin{document}
\label{firstpage}
\pagerange{\pageref{firstpage}--\pageref{lastpage}}
\maketitle

% Abstract of the paper
\begin{abstract}

This study presents the results concerning six red giant stars members of the globular cluster NGC~6558. Our analysis utilized high-resolution near-infrared spectra obtained through the CAPOS initiative (the APOgee Survey of Clusters in the Galactic Bulge), which focuses on surveying clusters within the Galactic Bulge, as a component of the Apache Point Observatory Galactic Evolution Experiment II survey (APOGEE-2). We employ the \texttt{BACCHUS} (Brussels Automatic Code for Characterizing High accUracy Spectra) code to provide line-by-line elemental-abundances for Fe-peak (Fe, Ni), $\alpha$-(O, Mg, Si, Ca, Ti), light-(C, N), odd-Z (Al), and the $s$-process element (Ce) for the 4 stars with high signal-to-noise ratios. This is the first reliable measure of the CNO abundances for NGC~6558.  Our analysis yields a mean metallicity for NGC~6558 of $\langle$[Fe/H]$\rangle$ = $-1.15 \pm 0.08$, with no evidence for a metallicity spread. We find a Solar Ni abundance, $\langle$[Ni/Fe]$\rangle$ $\sim$ $+$0.01, and a moderate enhancement of $\alpha$-elements, ranging between $+$0.16 to $<+$0.42, and a slight enhancement of the $s$-process element  $\langle$[Ce/Fe]$\rangle$ $\sim$ $+$0.19. We also found low levels of $\langle$[Al/Fe]$\rangle \sim +0.09$, but with a strong enrichment of nitrogen, [N/Fe]$>+0.99$, along with a low level of carbon, [C/Fe]$<-0.12$. This behaviour of Nitrogen-Carbon is a typical chemical signature for the presence of multiple stellar populations in virtually all GCs; this is the first time that it is reported in NGC~6558. We also observed a remarkable consistency in the behaviour of all the chemical species compared to the other CAPOS bulge GCs of the same metallicity.
\end{abstract}

% Select between one and six entries from the list of approved keywords.
% Don't make up new ones.

%%%% INTRODUCTION %%%%%%%%%%%%%%%%%%%%%%%%%%%%%%%%%%%%%       

\section{Introduction}
\label{section1}

The Milky Way bulge is a centrally concentrated component that has an associated system of globular clusters \citep{minniti1995}. 
Globular clusters (GCs) in the central region \citep[$\lesssim$ 3.5 kpc;][]{barbuy2018chemodynamical, Barbuy2018} of the Galactic Bulge (GB) could contain the oldest stars in the Galaxy.  Indeed, the rapid chemical evolution in the deep potential well of the GB means that GCs could be potentially older than their halo cousins, which were born and raised in much less massive satellites \citep{Cescutti2008, Barbuy2018, Savino2020, Geisler2021}. Thus, the bulge GCs are an excellent fossil to better understand the assembly of the most ancient part of our Galaxy. They are believed to have developed \textit{in situ} \citep{Massari2019}, as opposed to the halo GCs, all or mostly of which originated in satellite galaxies that later merged with the proto-Milky Way.

NGC~6558 is an important member of these ancient bulge GCs (BGCs). This cluster presents a post-core collapse structure \citep{Trager1995}, and lies in a relatively clear region inside the bulge (\textit{l} = $+$0.20096$^{\circ}$,  \textit{b} = $-$6.02488$^{\circ}$), with a mean heliocentric distance of $\sim$7.47$^{+0.29}_{-0.28}$ kpc \citep{2021-Baumgardt-Vasiliev, moreno2022} and a mean Galactocentric distance of $\sim$1.3 kpc. In an early work, \cite{rich1998} used $VI$ photometry to identify a strong blue extended horizontal branch in NGC~6558 with a metallicity estimated to lie in the range of $-$1.6 <$\langle$[Fe/H]$\rangle$<$-$1.2. Most recently, \cite{cohen2021} resolved the cluster main-sequence turnoff using the filters F606W and F814W from the Hubble Space Telescope (HST), and determined an age of $\sim$12.3 $\pm$ 1.1 Gyr. For their part, \cite{barbuy2007,Barbuy2018} presented for the first time a detailed spectroscopic analysis of five giant stars in NGC~6558 using high-resolution GIRAFFE optical spectra ($R\sim$ 22.000), and four giant stars using very high-resolution FLAMES-UVES spectra ($R\sim$ 45.000). These authors obtained precise abundances of the light elements C, N, O, the odd-Z elements Na, Al, the $\alpha$-elements Mg, Si, Ca, Ti, and the heavy elements Y, Ba, La, and Eu. They derived a mean metallicity of $\langle$[Fe/H]$\rangle$ = $-$1.17 $\pm$ 0.10 for the cluster via UVES spectra, with strong enhancements of $+$0.40 and $+$0.33 in the  $\alpha$-elements O and Mg, but lower enhancements of $+$0.11 and $+$0.07 in Si and Ca, respectively. The Odd-Z element Na showed a Solar value, while Al was slightly enhanced, with  $\langle$[Al/Fe]$\rangle$=$+$0.15, suggesting that these elemental abundances are in good agreement with the typical abundance behaviour found in other old BGCs such as NGC~6522 \citep{Barbuy2009, Barbuy2014, Trincado2019} and HP~1 \citep{Geisler2021}. On the other hand, the authors also noted a spread in the abundances of Y and Ba, compatible with expectations from massive spinstars \citep{chiappini2011,frischknecht2016}, while the $r$-element Eu is highly enhanced, with $\langle$[Eu/Fe]$\rangle$=$+$0.6, compatible with production by supernovae type II (SNII) or neutron star mergers \citep{goriely2011,wanajo2021}. Such high values of Eu are usually found in field halo stars, while higher enhancements are typically found in some ultra-faint dwarf galaxies \citep[see, e.g.,][]{ji2016,roederer2016,hansen2017,simon2019,reichert2020}. 

Kinematically, \cite{barbuy2007,Barbuy2018}, obtained precise mean radial velocities (RVs) for the cluster of $-$197.3$\pm$4.0 km\,s$^{-1}$ and $-$194.45$\pm$2.1 km\,s$^{-1}$, respectively. With the close proximity of NGC~6558 to the GB, along with its proper motion, \cite{rossi2015} and \cite{Perez-Villegas2020} found that NGC~6558 is probably an \textit{in situ} formed GC, co-rotating with the bar, similar to other GCs they classified. Recent works like \cite{Massari2019} and \cite{malhan2022} found similar results using the proper motions and parallaxes from the \textit{Gaia} releases available at that time.

Nowadays, with the high-precision astrometric and photometric data from the latest release of the ESA \textit{Gaia} mission \citep[\textit{Gaia} DR3;][]{GaiaDR3}, along with the photometric depth in both the Two-Micron All Sky Survey \citep{skrutskie2006} and the infrared survey VISTA Variables in the V\'ia L\'actea \citep[VVV,][]{minniti2010,Saito2012}, as well as the accurate chemical elements provided by the seventeenth data release of the APOGEE-2 survey \citep{Majewski2017, Accetta2022}, there is an unprecedented opportunity to examine in great detail these ancient GCs toward the GB, despite their location within very crowded and highly reddened regions. This is indeed the goal of the CAPOS project \citep{Geisler2021}.

In this paper, we present $H$-band spectroscopy for six members of NGC~6558, using spectra from the bulge Cluster APOgee Survey \citep[CAPOS,][]{Geisler2021}, an internal program of the Apache Point Observatory Galactic Evolution Experimental II survey (APOGEE-2, \citealt{Majewski2017}) and the Sloan Digital Sky IV survey \citep{Blanton2017, Ahumada2020}. These spectra make it possible to derive detailed elemental abundances for a large number of giant stars, as well as to achieve RVs with precision better than 1 km\,s$^{-1}$. 

This paper is organized as follows: Section \ref{section2} and  \ref{section3} present an overview of the CAPOS data analysed toward the NGC~6558 field, while Section \ref{section4} describes the adopted atmospheric parameters employed to determine the elemental abundances. The resulting elemental abundances obtained with the \texttt{BACCHUS} code are described in Section \ref{section5}. We discuss our results and compare them with previous literature in Section \ref{section6}. Our conclusions are presented in Section \ref{section7}.

\begin{table*}
	\setlength{\tabcolsep}{1.0mm}  
	\centering
	\caption{APOGEE Identifications, \textit{Gaia} DR3 PMs, \textit{Gaia} DR3 and VVV magnitudes, radial velocities, spectral signal-to-noise ratio (S/N), and APOGEE-2S programs names of targets toward NGC~6558.}
	\label{table1}
		\begin{tabular}{lccccccccccc}
			\hline 
			\multicolumn{1}{c}{APOGEE ID} & $\mu_{\alpha}\cos(\delta)$ & $\mu_{\delta}$ & G     & BP    & RP    & J    & H & K$_{\rm s}$   & RV          & S/N  & APOGEE-2  \\ 
			& (mas yr$^{-1}$)            & (mas yr$^{-1}$)              &         &         &         &        &        &        & (km s$^{-1}$) & (pixel$^{-1}$)   &  programs\\
			\hline
			High S/N   &      &      &  &  & &  & & &      & & \\
			\hline
			2M18101354$-$3146216            & $-1.98$      & $-4.04$      & 14.15 & 14.98 & 13.22 & 11.87 & 11.11 & 10.94 & $-$191.7  & 160  & \texttt{geisler\_19b}\\
			AP18102342$-$3146515            & $-1.69$      & $-4.11$      & 14.41 & 15.22 & 13.48 & 12.16 & 11.41 & 11.26 & $-$193.26 & 130  & \texttt{geisler\_19b} \\
			2M18101117$-$3144580            & $-1.70$      & $-4.24$      & 13.90 & 14.53 & 12.80 & 11.60 & 10.81 & 10.61 & $-$194.13 & 103  & \texttt{geisler\_19b} \\
			2M18101768$-$3145246            & $-1.94$      & $-3.94$      & 15.10 & 15.75 & 14.13 & 12.97 & 12.31 & 12.18 & $-$190.64 & 80   & \texttt{geisler\_19b} \\
			\hline
			Low S/N   &      &      &  &  & &  & & &      & & \\
			\hline
			2M18102223$-$3145435            & $-1.69$     & $-4.17$      & 15.38 & 16.11 & 14.52 & 13.28 & 12.59 & 12.47 & $-$193.44   & 46   & \texttt{geisler\_19b}\\
			2M18101623$-$3145423            & $-1.77$    & $-4.10$       & 15.98 & 16.30 & 15.18 & 14.51 & 14.13 & 14.03 & $-197.22$   & 10   & \texttt{kollmeier\_17a}\\
			\hline
		\end{tabular}
\end{table*}

\begin{figure*}
	\begin{center}
		\includegraphics[height = 14.cm]{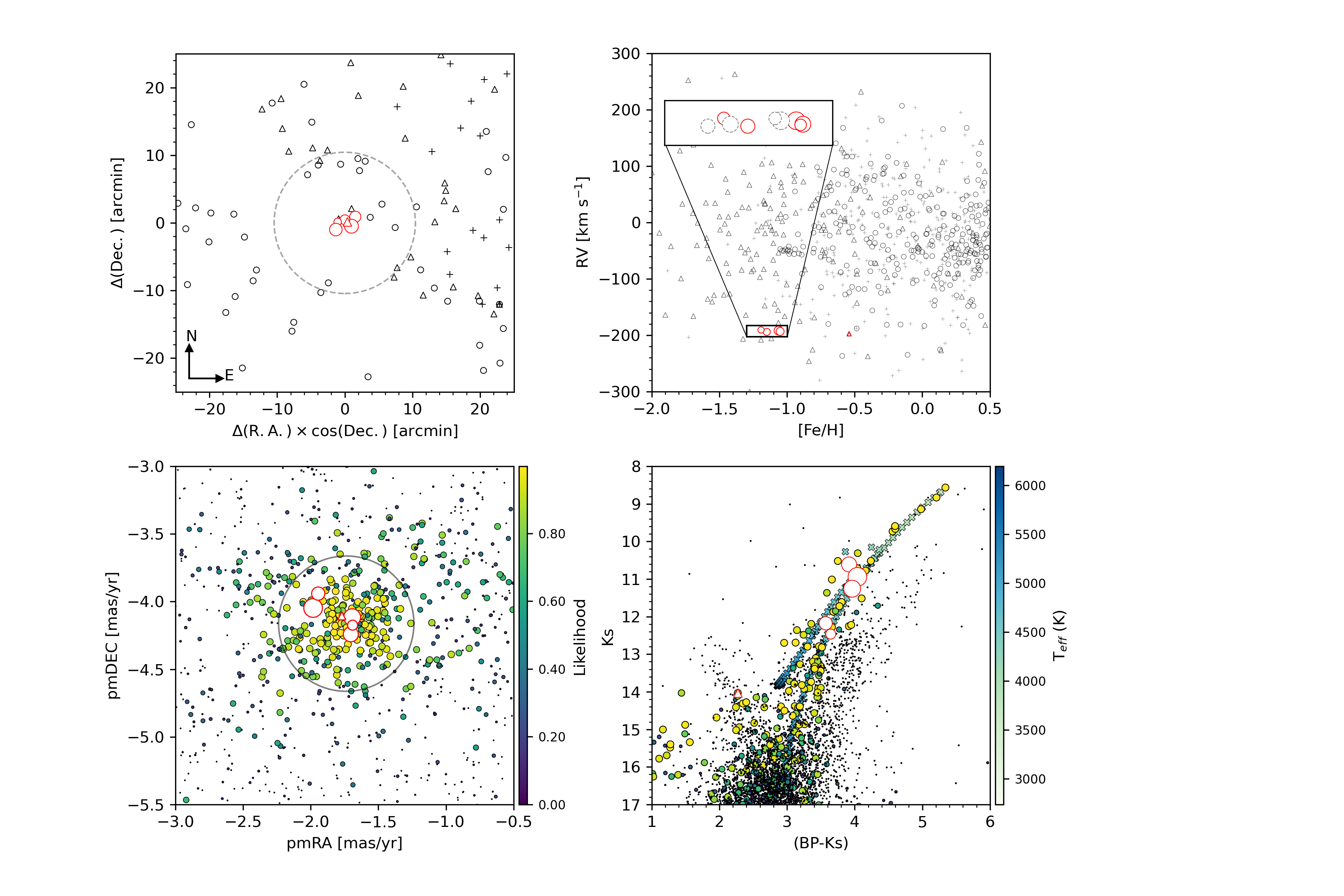}\\
		\caption{Distribution of the stars toward the NGC~6558 field. In all panels, the six cluster stars are identified with red open symbols. The size of the symbols is proportional to  the signal-to-noise ratio (S/N). The size of the smaller S/N (red triangle symbol star) was scaled by a factor of 3 in order to make it visible. \textit{Top-left panel}: spatial distribution of stars within the APOGEE-2 survey area. The black symbols refer to the overlapping APOGEE-2S plug-plates toward NGC~6558, corresponding to three APOGEE-2 programs: \texttt{geisler\_19b} (open circles), \texttt{kollmeier\_17a} (open triangles), and  \texttt{apogee} (plus symbols). The cluster tidal radius \citep[ r$_{\rm t} = $10\farcm44 ; ][]{Harris1996, Harris2010} is marked with a large open black dashed circle. North is up and east is to the right. \textit{Top-right panel}: plot of the radial velocity against metallicity of our member stars with the APOGEE-2S field sources in the direction of NGC~6558. Both parameters are from the ASPCAP pipeline. The black box, limited by $\pm$0.15 dex and $\pm$15 km\,s$^{-1}$ and centered on [Fe/H] = $-$1.15 and RV = $-$192.63 km\,s$^{-1}$, encloses our potential cluster members (red open circles), while the final sample of our four members whose abundances were determined with the BACCHUS code are the open dashed circles within the zoomed box (see text). \textit{Bottom-left panel}: proper motion distribution of the stars within the cluster tidal radius of NGC~6558 extracted from \textit{Gaia} DR3. The colour-coded bar represents the membership probabilities.  The probability of membership of our six stars is greater than 95\%. A circumference with a radius of 0.5 mas\,yr$^{-1}$ is shown in the proper motions diagram. \textit{Bottom-right panel}: $Ks$ vs. ($BP-Ks$) colour-magnitude diagram.  An isochrone with colour-coded cross-symbol temperatures has been fit to the most likely cluster members.}
		\label{figure1}
	\end{center}
\end{figure*}

%
%                                                           Section
%%%%%%%%%%%%%%%%%%%%%%%%%%%%%%%%%%%%%%%%%%%%%%%%%%%%%%%%%%% Data
\section{Observations and Data}
\label{section2}

The second phase of the Apache Point Observatory Galactic Evolution Experiment \citep[APOGEE-2;][]{Majewski2017} is a high-resolution ($R\sim22,500$) near-infrared (NIR) spectroscopic survey containing observations of $\sim657,135$ unique stars released as part of the SDSS-IV survey \citep{Blanton2017}, targeting these objects with selections detailed in \citet{Zasowski2013}, \citet{Zasowski2017}, \citet{Beaton2021}, and \citet{Santana2021}. These papers additionally provide an extensive explanation of the targeting approach employed in the APOGEE-2 survey. The spectra were obtained using the cryogenic, multi-fiber (300) APOGEE spectrograph \citep{Wilson2019}, mounted on the 2.5m SDSS telescope \citep{Gunn2006} at Apache Point Observatory, to observe the Northern Hemisphere (APOGEE-2N), and expanded to include a second APOGEE spectrograph on the 2.5m Ir\'en\'ee du Pont telescope \citep{Bowen1973} at Las Campanas Observatory to observe the Southern Hemisphere (APOGEE-2S). Each instrument registers the \textit{H}-band in wavelengths between the $15100$ \AA\, and $17000$ \AA, utilizing three detectors. There are small gaps in coverage between them of about 100 \AA\, around $\sim$15800 \AA\, and $\sim$16400 \AA\,, respectively. Additionally, each fiber provides an on-sky field of view with an diameter of $\sim$2$\arcsec$ in the Northern instrument and 1$\farcs$3 in the Southern instrument.

The final version of the APOGEE-2 catalog was published in December 2021 as part of the 17$^{\rm th}$ data release of the Sloan Digital Sky Survey \citep[DR17;][]{Abdurro2022} and is available publicly online through the SDSS Science Archive Server and Catalog Archive Server\footnote{SDSS DR17 data: \url{https://www.sdss4.org/dr17/irspec/spectro\_data/}}. The spectra were reduced following the procedures extensively described in \citet{Nidever2015}. For detailed information, we refer the interested reader to that reference. The stellar parameters and chemical abundances in APOGEE-2 were derived within the APOGEE Stellar Parameters and Chemical Abundances Pipeline \cite[\texttt{ASPCAP};][]{Garcia2016}. \texttt{ASPCAP} derives stellar-atmospheric parameters, radial velocities, and as many as 20 individual elemental abundances for each APOGEE-2 spectrum by comparing each to a multidimensional grid of theoretical \texttt{MARCS} model atmospheres grid \citep{Zamora2015}, employing a $\chi^2$ minimization routine with the code \texttt{FERRE} \citep{Allende2006} to derive the best-fit parameters for each spectrum. 

The accuracy and precision of the atmospheric parameters and chemical abundances are extensively analysed in \citet{Holtzman2018}, \citet{Henrik2018}, and \citet{Henrik2020}, while details regarding the customized \textit{H}-band line list are fully described in \citet{Shetrone2015}, 
\citet{Hasselquist2016}, \citet{Cunha2017}, and \citet{Smith2021}.

%
%                                                           Section
%%%%%%%%%%%%%%%%%%%%%%%%%%%%%%%%%%%%%%%%%%%%%%%%%%%%%%%%%%% NGC6558
\section{NGC~6558}
\label{section3}

NGC~6558 was observed as part of the Contributed APOGEE-2S CNTAC\footnote{CNTAC: Chilean Telescope Allocation Committee} CN2019B program (P.I: Doug Geisler; \texttt{geisler\_19b}) during July 9–10, 2019, as part of the CAPOS survey \citep{Geisler2021}.

Fig. \ref{figure1} shows the spatial distribution of the stars toward the NGC~6558 field in the APOGEE-2S footprint (top-left panel), together with the six identified cluster members listed in this work. The main APOGEE-2S plug-plate containing the NGC~6558 cluster (toward $\alpha_{\mathrm{J2000}}$ $=$ 18$^{\rm h}$10$^{\rm m}$18.38$^{\rm s}$ and $\delta_{\mathrm{J2000}}$ = $-$31$^{\rm d}$45$^{\rm m}$48.6$^{\rm s}$) was centered on (\textit{l,b}) $\sim$ (0.0, $-$0.6) degrees, and 15 out of 550 science fibers were located inside the cluster tidal radius \citep[r$_t = 10\farcm44$ ;][ dashed circumference in Fig. \ref{figure1}]{Harris1996, Harris2010}. 

CAPOS targets were originally selected on the basis of \textit{Gaia} DR2 \citep{GaiaDR2} to have proper motions (PMs) within an approximate radius of $\sim$0.5 mas yr$^{-1}$ around the nominal PMs of NGC~6558; $\mu_{\alpha}\cos{}$($\delta$)$=-1.72$ and $\mu_{\delta} = -4.14$ \citep[see, e.g.][]{Baumgardt-Vasiliev2021}. We initially selected these values to estimate candidate members from the PMs to reduce the potential field star contamination in both optical and NIR color-magnitude diagrams (CMDs). We re-examined the PMs from the original sample, but we did not find important differences between  \textit{Gaia} DR2 \citep{GaiaDR2} and \textit{Gaia} DR3 PMs \citep{GaiaDR3} toward NGC~6558.  The left-bottom panel in Fig. \ref{figure1} shows the updated version of the PMs using \textit{Gaia} DR3\textit{Gaia} \citep{GaiaDR3}. Besides PMs, the selected stars are distributed along both the red giant branch (RGB) and the asymptotic giant branch (AGB) of the cluster, as illustrated in the bottom-right panel of Fig. \ref{figure1}.

Our target stars not only lie inside the tidal radius of the cluster, but also share similar PMs, follow the same evolutionary track in the optical$+$NIR CMDs, and share similar radial velocity (RV) and metallicity ([Fe/H], top-right panel), which confirm that the 5 CAPOS stars (the red open circles) are very likely cluster members. Both the RV and [Fe/H] were extracted from the spectra reduced with the \texttt{ASPCAP} pipeline, and these values in the metallicity were used only for the member selection. In fact, \cite{BacchuAspcap2020} shows that there is usually an offset of $\sim$0.2\,dex on average between the abundances measured with \texttt{ASPCAP} pipeline with respect to the BACCHUS code in most of the chemical species. The methodology followed to determine the final values of the abundances is described in detail in Sections \ref{section4} and \ref{section5}. In this sense, the zoomed black box rectangle in the top-right panel of the Fig. \ref{figure1} shows the revised Fe abundances as black dashed open circles, together with the \texttt{ASPCAP} values (red open circles). The APOGEE-2 designations of the stars, the coordinates, \textit{Gaia} DR3 and VVV magnitudes, RV, and S/N are listed in Table \ref{table1}.

Additionally, we also examined two overlapping APOGEE-2S plug-plates toward NGC~6558, which contain the APOGEE-2S programs: \texttt{kollmeier\_17a} (238 science fibers) and  \texttt{apogee} (332 science fibers). We refer the interested reader to \citet{Santana2021} for further details on these programs. 

From  \texttt{kollmeier\_17a}, we have identified one star (2M18101623$-$3145423) satisfying almost all the cluster membership criteria imposed in this work. However, this star is only recorded in Table \ref{table1} together with the APOGEE-2 RV, \textit{Gaia} DR3 astrometry, and photometry information because its spectral S/N is too low ($<10$) to obtain reliable elemental abundances. This star is positioned along the cluster CMD's horizontal branch (HB), and is likely associated with the RR Lyrae population of NGC~6558.

Finally, we imposed a quality cut in the APOGEE-2 spectra using an S/N threshold $>$ 70 per pixel to ensure the analysis of good-quality spectra and maximize the precision of the stellar abundances. Thus, elemental abundances are restricted to the 4 stars with high S/N listed in Table \ref{table1}.

\begin{table}
	\begin{center}
	\setlength{\tabcolsep}{3.0mm}  
	\caption{Estimated photometric atmospheric parameters for  the NGC~6558 stars with high S/N (see text).}
	\label{table2}
		\begin{tabular}{lccc}
			\hline
			APOGEE ID  & T$_{\mathrm{eff}}$ & $\log$ \textit{g}  &  ${\xi}_t$  \\
			& (K) & (c.g.s)   & (km\,s$^{-1}$)  \\
			\hline
			2M18101354$-$3146216 & 4319      & 1.3      & 1.61     \\
			AP18102342$-$3146515 & 4305      & 1.4       & 1.48     \\			
			2M18101117$-$3144580 & 4381      & 1.1       & 1.80     \\
			2M18101768$-$3145246 & 4729      & 1.8      & 1.61       \\
			\hline
		\end{tabular}
  \end{center}
\end{table}

\begin{figure}
	\begin{center}
		\includegraphics[trim = 20mm 20mm 20mm 25mm, clip, angle=0, width=8.5cm]{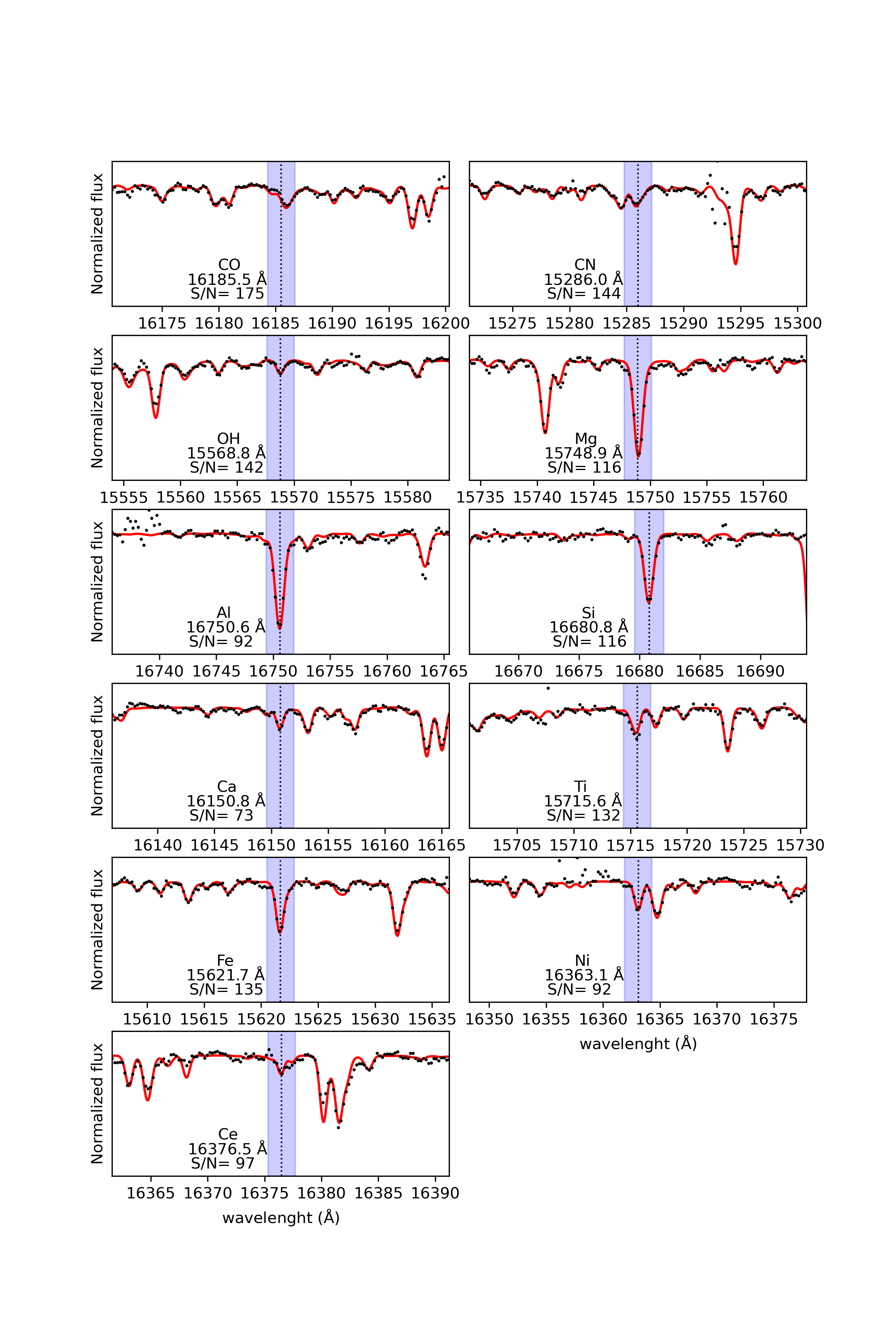}\\
		\caption{Example of the quality of the model fit obtained with \texttt{BACCHUS} in this study for selected molecular lines: $^{12}$C$^{16}$O, $^{12}$C$^{14}$N, $^{16}$OH, and atomic lines: Mg~I, Al~I, Si~I, Ca~I, Ti~I, Fe~I, Ni~I, and Ce~II for one target giant in NGC~6558: AP18102342$-$3146515. The best-fit synthesis and observed spectra are highlighted with red lines and black dots, respectively. Each panel is centered around the selected lines, and the dashed black lines mark the position of the air wavelength lines.}
		\label{figure2}
	\end{center}
\end{figure}

\begin{table*}
\begin{center}
    \setlength{\tabcolsep}{1.0mm}
    \caption{Elemental abundances of NGC~6558 CAPOS stars. Within parentheses are the re-calibrated abundances for the star in common with \protect\cite{Barbuy2018} (2M18101768$-$3145246).}
    \label{tab:abundances}
    
    \begin{tabular}{cccccccccccc}
        \hline
        ID & {[}C/Fe{]} & {[}N/Fe{]} & {[}O/Fe{]} & {[}Mg/Fe{]} & {[}Al/Fe{]} & {[}Si/Fe{]} & {[}Ca/Fe{]} & {[}Ti/Fe{]} & {[}Fe/H{]} & {[}Ni/Fe{]} & {[}Ce/Fe{]} \\
        \hline
        2M18101354$-$3146216 & $-$0.42 & $+$1.31 & $+$0.36   & $+$0.24   & $+$0.05   & $+$0.21   & $+$0.31  & $+$0.16  & $-$1.09   & $-$0.07 & 0.18 \\
        AP18102342$-$3146515 & $-$0.40 & $+$1.13 & $+$0.29   & $+$0.23   & $+$0.29   & $+$0.22   & $+$0.35  & $+$0.20  &$-$1.18    & $-$0.10 & 0.13 \\
        2M18101117$-$3144580 & $-$0.12 & $+$0.99 & $+$0.37   & $+$0.22   & $-$0.04   & $+$0.27   & $+$0.28  & $+$0.13  &$-$1.22    & $+$0.14 & 0.25 \\
        2M18101768$-$3145246 & ...     & ...     & $+$0.63   & $+$0.22   & $+$0.08   & $+$0.25   & $+$0.30  & $+$0.14  &$-$1.10    & $+$0.08 & ... \\
                             & ...     & ...     & ($+$0.62) & ($+$0.40) & ($+$0.05) & ($+$0.19) & ($+$0.16) &($+$0.16) & ($-$1.10) & ...    & ...  \\
        \hline
        \multicolumn{1}{c}{Mean} & $-$0.31 & $+$1.14 & $+$0.41 & $+$0.23 & $+$0.09 & $+$0.24 & $+$0.31 & $+$0.16 & $-$1.15 & $+$0.01 & $+$0.19 \\
        \hline
    
    \end{tabular}
\end{center}
\end{table*}

%
%                                                           Section
%%%%%%%%%%%%%%%%%%%%%%%%%%%%%%%%%%%%%%%%%%%%%%%%%%%%%%%%%%% Atmospheric
\section{Atmospheric Parameters}
\label{section4}

The atmospheric parameters (T$_{\rm eff}$ and $\log g$) were derived using the same methodology as described in \citet{Fernandez-Trincado2022}, that is, the $Ks$ vs. $BP-Ks$ CMD shown in Fig. \ref{figure1} (bottom-right panel) was reddening corrected by using giant stars and adopting the reddening law of \citet{Cardelli1989} and \citet{Donnell1994} and a total-to-selective absorption ratio RV $=$ 3.1 (see, \citealp{Barbuy2018,cohen2021}). 

We chose the RGB stars located within a 10$\arcmin$ radius from the cluster, ensuring that their proper motions closely matched those of NGC~6558 within a margin of 0.5~mas\,yr$^{-1}$. Subsequently, following the reddening slope, we proceeded to determine the separation between these selected stars and a previously defined reference line, which outlines the RGB. The vertical and horizontal separation yields the differential absorption A$_{\rm K_s}$ and reddening E(BP $-$ K$_{\rm s}$) values of each individual star, respectively. Our next step involved the selection of the three closest RGB stars within the field to calculate the average absorption and reddening values. We then corrected the K$_{\rm s}$ magnitude and  BP $-$ K$_{\rm s}$ colour index based on these averages, respectively. The selection of this number of stars represents an arbitrary decision aimed at minimizing the impact of photometric random errors while maximizing spatial resolution to the greatest extent possible.  After these corrections the photometric $T_{\rm eff}$ and $\log$ \textit{g} were determined by fitting a \texttt{PARSEC} isochrone \citep{Bressan2012}, chosen to have an age of 12.3 Gyr \citep{cohen2021}, and an overall metallicity of [M/H]=$-$0.90 (this work), and assumed $T_{\mathrm{eff}}$ and $\log$ \textit{g}  to be the effective temperature and gravity at the point of the isochrones that has the same K$_{\rm s}$ magnitude as the star. Finally, microturbulence velocities $\xi_t$ were determined using the relation from \cite{gratton1996}. Table \ref{table2} lists the determined photometric atmospheric parameters, microturbulence velocity, and S/N for the four stars in our sample for which elemental abundances are provided in this study.

%
%                                                           Figure
%%%%%%%%%%%%%%%%%%%%%%%%%%%%%%%%%%%%%%%%%%%%%%%%%%%%%%%%%%% Abundances
\begin{figure*}
\begin{center}
\includegraphics[height = 10.cm]{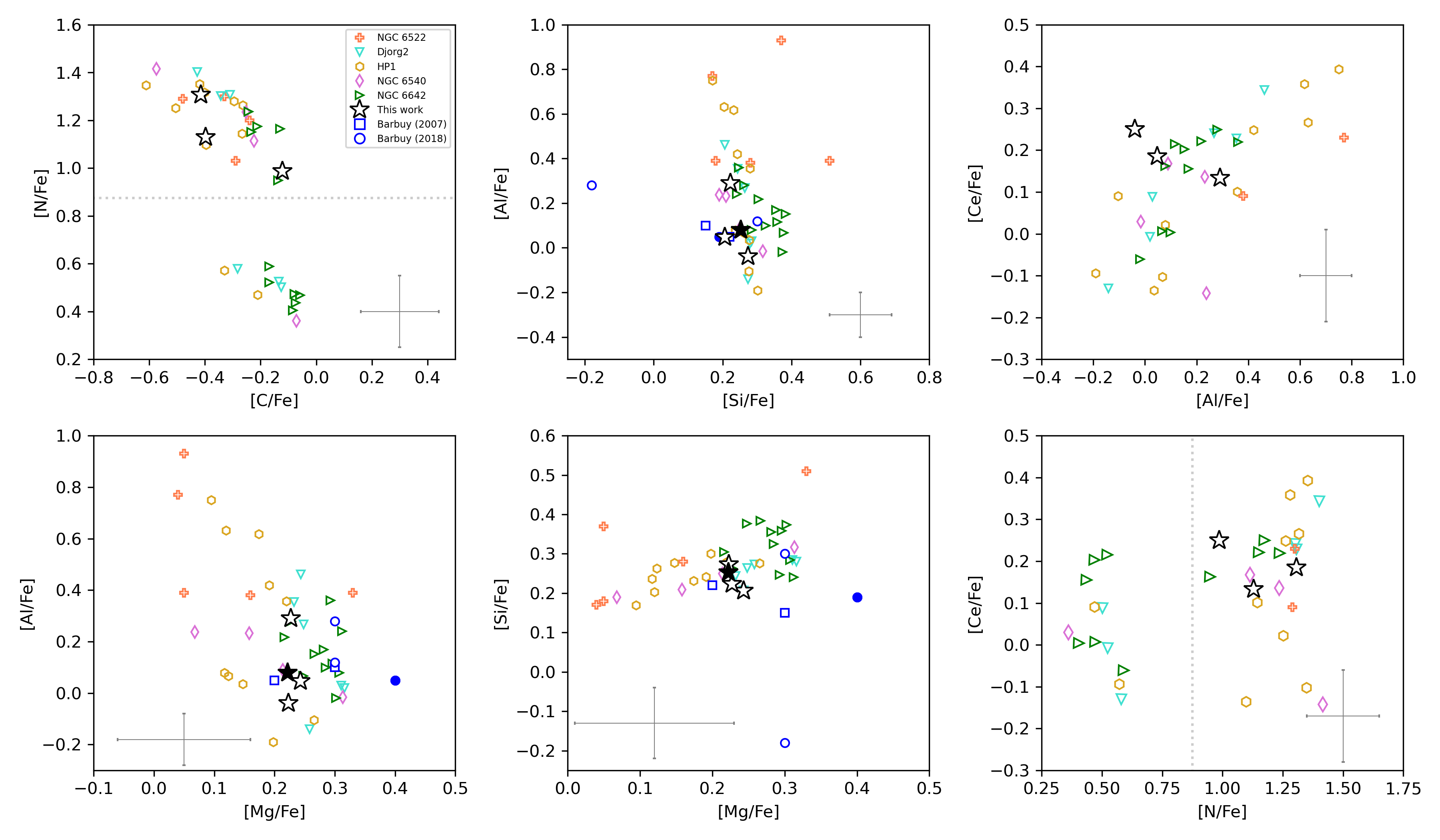}\\
\caption{[N/Fe] vs [C/Fe], [Al/Fe] vs [Si/Fe], [Ce/Fe] vs [Al/Fe], [Al/Fe] vs [Mg/Fe], [Si/Fe] vs [Mg/Fe], and [Ce/Fe] vs [N/Fe] diagrams for our NGC~6558 stars (black open star symbols). Open blue squares and circles depict the stars analysed by \protect\cite{barbuy2007} and  \protect\cite{Barbuy2018} at the same cluster, respectively. Filled symbols represent the star in common with our study. The figure also displays data for the bulge cluster NGC~6522 \protect\citep[open plus symbols,][]{Fernandez-Trincado2019NGC6522}, as well as the CAPOS clusters Djorg~2, HP~1, NGC~6540, and NGC~6642 \citep{Geisler2021}. Grey dotted lines mark the level of 4$\sigma$ above the Galactic level for [N/Fe] enhanced stars at the metallicity of NGC~6558 (see Fig. \ref{fig:Polyplot}).}
\label{figure3}
\end{center}
\end{figure*}
%
%                                                           Figure
%%%%%%%%%%%%%%%%%%%%%%%%%%%%%%%%%%%%%%%%%%%%%%%%%%%%%%%%%%% END

%                                                           Figure
%%%%%%%%%%%%%%%%%%%%%%%%%%%%%%%%%%%%%%%%%%%%%%%%%%%%%%%%%%% Abundances
\begin{figure}
\begin{center}
\includegraphics[height = 8.5cm]{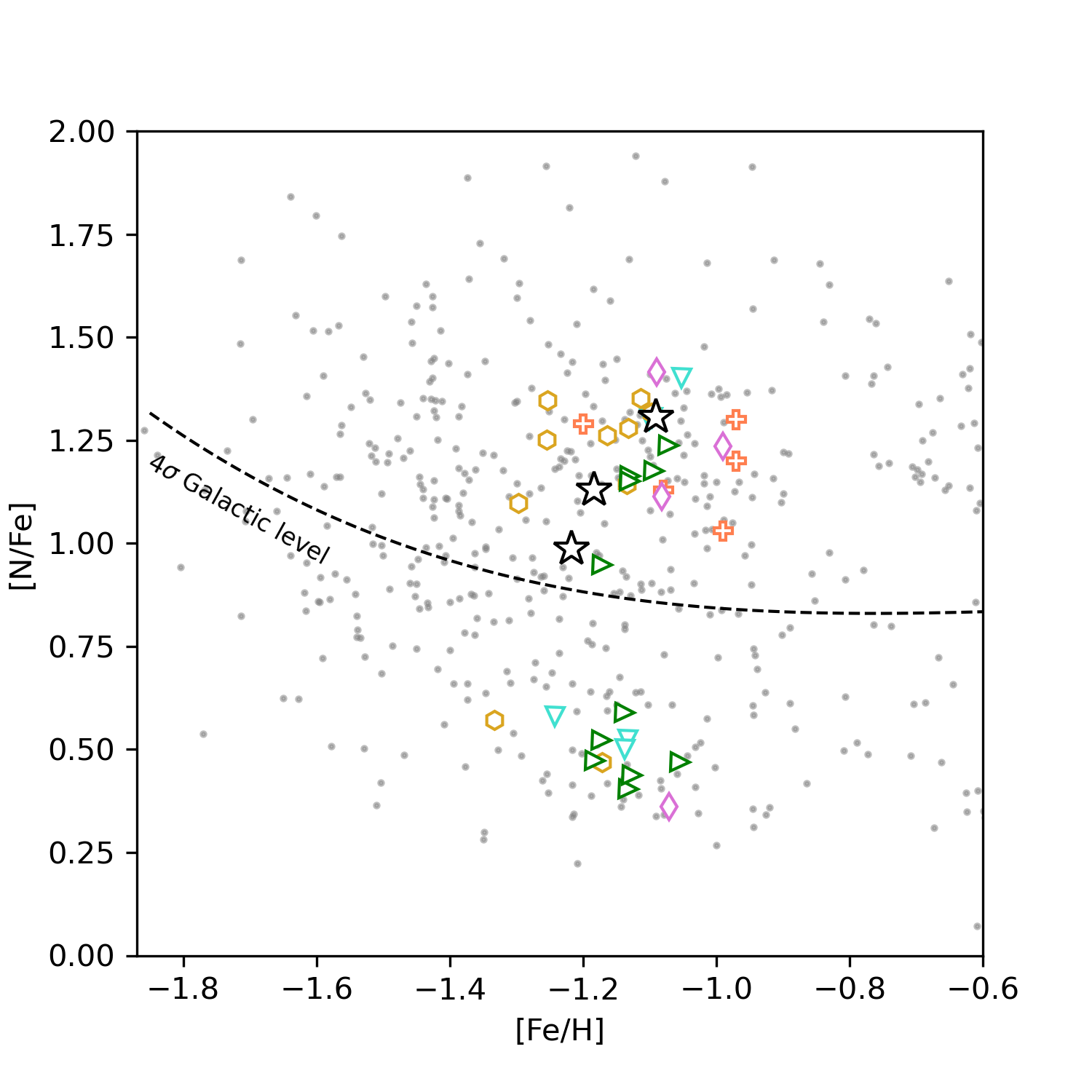}\\
\caption{ Distribution of [N/Fe] vs. [Fe/H] for the stars in the GCs from \citet[][grey dots]{Meszaros2020}. Symbols and colour
codes are the same as in Fig. \ref{figure3}. The dashed black line represents a 4$^{th}$ order polynomial fit at the 4$\sigma$ level, over the usual Galactic values as described in \citet{fernandez2022galactic}.}
\label{fig:Polyplot}
\end{center}
\end{figure}
%
%                                                           Figure
%%%%%%%%%%%%%%%%%%%%%%%%%%%%%%%%%%%%%%%%%%%%%%%%%%%%%%%%%%% END

%
%                                                           Figure
%%%%%%%%%%%%%%%%%%%%%%%%%%%%%%%%%%%%%%%%%%%%%%%%%%%%%%%%%%% Violin
\begin{figure*}
\begin{center}
\includegraphics[width=14cm]{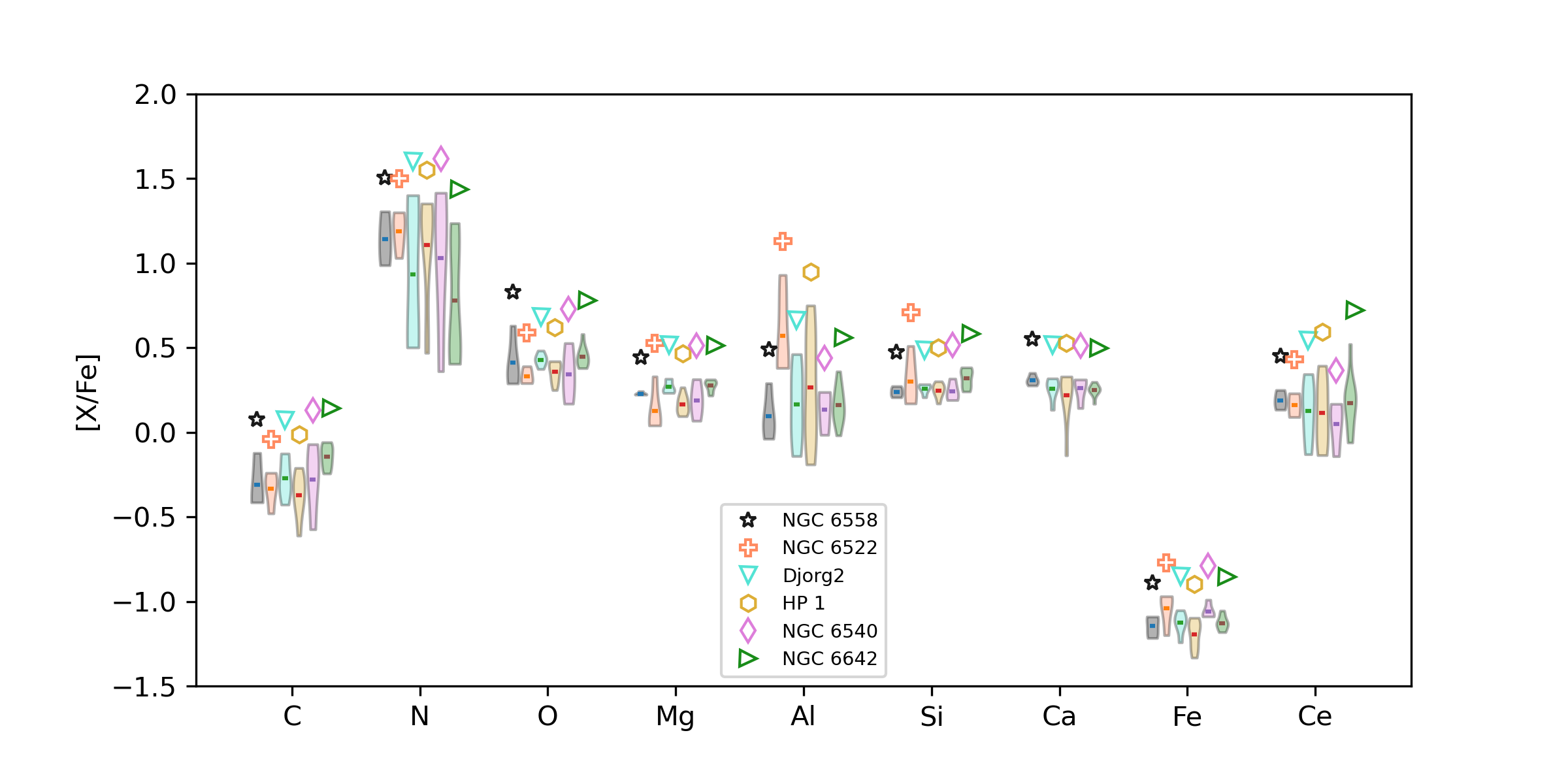}\\
\caption{Violin plot for the abundances of C, N, O, Mg, Al, Si, Ca, Fe, and Ce of NGC~6558 compared with other CAPOS BGCs.}
\label{figure4}
\end{center}
\end{figure*}
%
%                                                           Figure
%%%%%%%%%%%%%%%%%%%%%%%%%%%%%%%%%%%%%%%%%%%%%%%%%%%%%%%%%%% END

%
%                                                           Table
%%%%%%%%%%%%%%%%%%%%%%%%%%%%%%%%%%%%%%%%%%%%%%%%%%%%%%%%%%% Errors

\begin{table}
\normalsize
\caption{Typical abundance uncertainties determined for four stars in our sample.}
\label{tab:errores}
\begin{tabular}{lccccc}
\hline
\multicolumn{1}{c}{Abundance} & $\Delta T_{\mathrm{eff}}$ & $\Delta \log g$ & $\Delta \xi_t$ & Mean & ($\Sigma x^{2}$)$^{1/2}$ \\
\hline
  & (K) & (dex) & (km\,s$^{-1}$) & (dex) & (dex) \\
\hline
  &  &  &  &  &  \\
\hline
\multicolumn{3}{l}{2M18101354$-$3146216} & & & \\
\hline
{[}C/Fe{]} & 0.08 & 0.07 & 0.02 & 0.07 & 0.13 \\
{[}N/Fe{]} & 0.14 & 0.03 & 0.01 & 0.07 & 0.16 \\
{[}O/Fe{]} & 0.17 & 0.04 & 0.01 & 0.06 & 0.18 \\
{[}Mg/Fe{]} & 0.04 & 0.02 & 0.01 & 0.11 & 0.12 \\
{[}Al/Fe{]} & 0.09 & 0.01 & 0.01 & 0.07 & 0.12 \\
{[}Si/Fe{]} & 0.01 & 0.04 & 0.02 & 0.05 & 0.07 \\
{[}Ca/Fe{]} & 0.03 & 0.04 & 0.01 & 0.05 & 0.07 \\
{[}Ti/Fe{]} & 0.12 & 0.01 & 0.00 & 0.07 & 0.14 \\
{[}Fe/H{]} & 0.05 & 0.03 & 0.01 & 0.06 & 0.08 \\
{[}Ni/Fe{]} & 0.01 & 0.01 & 0.00 & 0.06 & 0.06 \\
 &  &  &  &  &  \\
\hline
\multicolumn{3}{l}{AP18102342$-$3146515}  & & & \\
\hline
{[}C/Fe{]} & 0.06 & 0.07 & 0.00 & 0.10 & 0.13 \\
{[}N/Fe{]} & 0.15 & 0.07 & 0.00 & 0.07 & 0.18 \\
{[}O/Fe{]} & 0.15 & 0.03 & 0.00 & 0.07 & 0.17 \\
{[}Mg/Fe{]} & 0.01 & 0.02 & 0.00 & 0.13 & 0.13 \\
{[}Al/Fe{]} & 0.04 & 0.01 & 0.00 & 0.08 & 0.09 \\
{[}Si/Fe{]} & 0.02 & 0.03 & 0.01 & 0.07 & 0.07 \\
{[}Ca/Fe{]} & 0.03 & 0.04 & 0.01 & 0.05 & 0.07 \\
{[}Ti/Fe{]} & 0.12 & 0.02 & 0.00 & 0.08 & 0.14 \\
{[}Fe/H{]} & 0.03 & 0.03 & 0.00 & 0.07 & 0.08 \\
{[}Ni/Fe{]} & 0.05 & 0.06 & 0.00 & 0.05 & 0.09 \\
 &  &  &  &  &  \\
\hline
\multicolumn{3}{l}{2M18101117$-$3144580} & & & \\
\hline
{[}C/Fe{]} & 0.03 & 0.06 & 0.02 & 0.07 & 0.10 \\
{[}N/Fe{]} & 0.11 & 0.02 & 0.00 & 0.07 & 0.13 \\
{[}O/Fe{]} & 0.19 & 0.02 & 0.01 & 0.06 & 0.19 \\
{[}Mg/Fe{]} & 0.03 & 0.01 & 0.00 & 0.11 & 0.12 \\
{[}Al/Fe{]} & 0.07 & 0.02 & 0.00 & 0.07 & 0.10 \\
{[}Si/Fe{]} & 0.02 & 0.05 & 0.01 & 0.05 & 0.07 \\
{[}Ca/Fe{]} & 0.03 & 0.04 & 0.00 & 0.05 & 0.08 \\
{[}Ti/Fe{]} & 0.13 & 0.02 & 0.00 & 0.07 & 0.15 \\
{[}Fe/H{]} & 0.05 & 0.04 & 0.00 & 0.06 & 0.08 \\
{[}Ni/Fe{]} & 0.00 & 0.01 & 0.00 & 0.06 & 0.06 \\
 &  &  &  &  &  \\
\hline
\multicolumn{3}{l}{2M18101768$-$3145246} & & & \\
\hline
{[}C/Fe{]} & ... & ... & ... & ... & ... \\
{[}N/Fe{]} & ... & ... & ... & ... & ... \\
{[}O/Fe{]} & 0.17 & 0.07 & 0.01 & 0.06 & 0.19 \\
{[}Mg/Fe{]} & 0.01 & 0.05 & 0.01 & 0.11 & 0.12 \\
{[}Al/Fe{]} & 0.02 & 0.02 & 0.00 & 0.07 & 0.08 \\
{[}Si/Fe{]} & 0.04 & 0.03 & 0.01 & 0.05 & 0.07 \\
{[}Ca/Fe{]} & 0.02 & 0.04 & 0.01 & 0.06 & 0.08 \\
{[}Ti/Fe{]} & 0.10 & 0.02 & 0.00 & 0.09 & 0.13 \\
{[}Fe/H{]} & 0.04 & 0.04 & 0.01 & 0.06 & 0.08 \\
{[}Ni/Fe{]} & 0.01 & 0.03 & 0.01 & 0.06 & 0.07 \\

\hline
\end{tabular}
\end{table}
%
%                                                           Table
%%%%%%%%%%%%%%%%%%%%%%%%%%%%%%%%%%%%%%%%%%%%%%%%%%%%%%%%%%% END

%
%                                                           Section
%%%%%%%%%%%%%%%%%%%%%%%%%%%%%%%%%%%%%%%%%%%%%%%%%%%%%%%%%%% Abundances
\section{Elemental Abundances}
\label{section5}

Elemental abundances were derived from a local thermodynamic equilibrium (LTE) analysis for the four stars listed in Table \ref{table2} using the Brussels Automatic Code for Characterizing High accuracy Spectra \citep{Masseron2016}, which relies on the radiative code \texttt{Turbospectrum} \citep{Alvarez1998, Plez2012} and the \texttt{MARCS} model atmosphere grid \cite{Gustafsson2008}. With the atmospheric parameters determined in Section \ref{section4}, the first step consisted in determining the metallicity ([Fe/H]) from selected Fe I lines and broadening parameters. 

With the [Fe/H] and main atmospheric parameters fixed, we then computed the elemental abundance of each chemical species following the same methodology as described in \citet{trincadoUKS1}, \citet{trincadoSgr}, \citet{trincadoVVVCL001}, and \citet{Fernandez-Trincado2022}, and summarized here for guidance. We performed a synthesis using the full set of atomic line lists described in \citet{Smith2021}, which is used to fit the local continuum.  Subsequently, the cosmic rays and the telluric lines are removed, and the local S/N is estimated. After that, we follow the description given in \cite{Hawkins16}. We used four different abundance determination methods, namely, line-profile fitting, core line-intensity comparison, global goodness-of-fit estimate, and an equivalent-width comparison. For each method, we used a set of synthetic spectra of different abundances to compute the observed abundance. Only the lines with the best fit of the individual abundances are kept. We chose the $\chi^{2}$ diagnostic for abundance determination due to its reliability. Nevertheless, we preserved data from the other diagnostic methods, including the standard deviation obtained from all four approaches to estimate the uncertainties. 

To determine the C, N, and O elemental abundances, a mix of heavily CN-cycled and $\alpha-$poor \texttt{MARCS} models were used, as well as the same molecular lines adopted by \cite{Smith2021}. The oxygen abundances are first estimated from the hydroxide molecular lines ($^{16}$OH). With this abundance known, we derived the carbon abundances from the carbon monoxide molecular lines ($^{12}$C$^{16}$O), and finally, the nitrogen abundances are obtained from the cyanogen molecular lines ($^{12}$C$^{14}$N). This process is repeated multiple times to eliminate any dependence on OH, CO, and CN lines.

An illustration of the quality of the model fit obtained in this study is presented in Fig. \ref{figure2}, where we show the observed (black dotted line) and synthetic (red line) spectra for the APOGEE-2 region around selected atomic and molecular lines obtained for our target giant star AP18102342$-$3146515.

Eleven chemical species were investigated for NGC~6558 from the APOGEE-2 spectra, including the Fe-peak (Fe, Ni), $\alpha-$(O, Mg, Si, Ca, Ti), light- (C, N), odd-Z (Al), and the $s$-process element (Ce). Our study provides a number of elements in common with optical studies carried out by \citet{barbuy2007} and \citet{Barbuy2018}, as well as complementary elements not previously examined. The resulting abundance derivations are listed in Table \ref{tab:abundances}.

Uncertainties in spectroscopic parameters are listed in Table \ref{tab:errores} for our four cluster stars. The reported uncertainties were obtained by once again calculating the abundances by varying  one by one the mean value of the effective temperature, surface gravity, and microturbulence velocity by $\Delta$T$_{\rm eff} = \pm 100$ K, $\Delta \log$ \textit{g} $= \pm 0.3$ dex,  $\Delta \xi_{t} = \pm 0.05$ km s$^{-1}$, respectively. These values represent the usual uncertainties for the stellar parameters in GCs \citep[see, e.g.,][]{Barbuy2018,Trincado2019}. The final uncertainty of each chemical abundance is the quadratic sum of the individual uncertainties: $\sigma^2_{\rm total}  = \sigma^2_{\rm [X/H], T_{eff}}    + \sigma^2_{\rm [X/H], \log g} + \sigma^2_{\rm [X/H],\xi_t}  + \sigma^2_{\rm mean} $; where $\sigma^2_{\rm mean}$ is the standard deviation of the abundances computed for each line of a given element. Table \ref{tab:errores} also shows the characteristic uncertainties for each abundance due to variations in the stellar parameters.

We derived a mean spectroscopic metallicity  $\langle$[Fe/H]$\rangle$ = $-1.15 \pm 0.08$ for NGC~6558, only 0.02 dex above of the results from \cite{Barbuy2018} ($\langle$[Fe/H]$\rangle$ = $-1.17$), who used UVES spectra. These authors investigated a star in common  with our data set, namely, the star 2M18101768$-$3145246, with [Fe/H] = $-$1.15 computed from both Fe I and Fe II lines. This value differs by 0.05 dex from ours (see Table \ref{tab:abundances}). \cite{Barbuy2018} adopted a Solar iron abundance of $\epsilon$(Fe)$_{\odot}$ = 7.50 from \cite{grevesse1998} for computing [X/Fe] abundances, in contrast to our study, which adopted that from \cite{Asplund2005}, $\epsilon$(Fe)$_{\odot}$= 7.45. In order to analyse the entire set of data available for NGC~6558, we re-calibrated the values of those authors to be at the same relative Solar iron abundances as this work. Re-calibrated values to Asplund's Solar abundances for the star in common with \cite{Barbuy2018} are listed in parentheses in Table \ref{tab:abundances}. In this way, the star in common now shows the same Fe abundance, and a mean value for GC NGC~6558 of $\langle$[Fe/H]$\rangle$ = $-$1.12 was obtained from \cite{Barbuy2018}. This value is in excellent agreement with our work.

We report for the first time for NGC~6558 abundances in the iron-peak nickel (Ni) and the $s$-process element cerium (Ce). The former shows an average abundance essentially Solar $\langle$[Ni/Fe]$\rangle$ = $+$0.01, and even deficient in two stars, with a clear separation of $\sim$ 0.2 dex, which is greater than their individual errors. The latter shows an average abundance slightly super-Solar, $\langle$[Ce/Fe]$\rangle$ = $+$0.19.

We also found strong enrichment for nitrogen, $\langle$[N/Fe]$\rangle$ $\sim +$1.14, well above typical Galactic field star levels \citep[see, e.g.,][]{fernandez2022galactic} for the 3 stars for which the spectra allowed us to measure CNO abundances, as well as carbon-deficient levels of $\langle$[C/Fe]$\rangle$ = $-$0.31. These high values for N agree with the average value of $\langle$[N/Fe]$\rangle$ = $+$1.01 reported by \cite{ernandes2018}, who analysed the same sample of stars  as in \cite{Barbuy2018}, but it strongly  disagrees with their average value for carbon, $\langle$[C/Fe]$\rangle$ = $+$0.23, which is 0.54 dex greater than our measures when their values are re-calibrated to \cite{Asplund2005}. Although our data use the $H$-band, where molecular lines permit us to achieve what we believe to be a more accurate determination of C, N, and O together, both the C and N are affected not only by stellar evolution but also by multiple populations. We did not obtain reliable measurements for C and N for our star in common, so a good agreement between the two samples is not necessarily expected.

The star 2M18101768$-$3145246 exhibits clear differences, about 0.18 dex lower in our measurement for [Mg/Fe] with  respect to \cite{Barbuy2018}, and 0.14 dex higher in [Ca/Fe], which fall outside our error bars, in contrast with the much smaller differences of only 0.01 and 0.03 dex for [O/Fe] and [Al/Fe] abundances, respectively, which together with Si and Ti are in agreement within the error bars.  

We found an average enhancement for the $\alpha$-element $\langle$[O/Fe]$\rangle$ of $+$0.41, and very similar but lower values for the $\alpha$-elements $\langle$[Mg/Fe]$\rangle$, $\langle$[Si/Fe]$\rangle$, and $\langle$[Ca/Fe]$\rangle$ of $+$0.23, $+$0.24, and $+$0.31 respectively. The O and Mg are thought to be produced in the hydrostatic phases of massive stars, while Si and Ca are formed in a SN type II \citep{heger2003,janka2012}.  In comparison, \cite{Barbuy2018} found high $\langle$[O/Fe]$\rangle$ and $\langle$[Mg/Fe]$\rangle$, but lower $\langle$[Ca/Fe]$\rangle$, $\langle$[Si/Fe]$\rangle$ and $\langle$[Ti/Fe]$\rangle$.

%
%                                                           Table
%%%%%%%%%%%%%%%%%%%%%%%%%%%%%%%%%%%%%%%%%%%%%%%%%%%%%%%%%%% OtherClus

\begin{table*}
\begin{center}
    \setlength{\tabcolsep}{1.0mm}
    \caption{\bf Mean abundances of NGC~6558 compared to selected CAPOS BGCs. }
    \label{tab:OtherClus}
    \begin{tabular}{lccccccccc}
    \hline

Cluster       & {[}C/Fe{]} & {[}N/Fe{]} & {[}O/Fe{]} & {[}Mg/Fe{]} & {[}Al/Fe{]} & {[}Si/Fe{]} & {[}Ca/Fe{]} & {[}Fe/H{]} & {[}Ce/Fe{]} \\
\hline

NGC~6558 & $-$0.31      & $+$1.14       & $+$0.41       & $+$0.23        & $+$0.09        & $+$0.24        & $+$0.31        & $-$1.15      & $+$0.19        \\
NGC~6522 & $-$0.34      & $+$1.19       & $+$0.33       & $+$0.13        & $+$0.57        & $+$0.30        & ---         & $-$1.04      & $+$0.16        \\
Djorg~2  & $-$0.27      & $+$0.94       & $+$0.43       & $+$0.27        & $+$0.17        & $+$0.26        & $+$0.26        & $-$1.13      & $+$0.13        \\
HP~1     & $-$0.37      & $+$1.11       & $+$0.36       & $+$0.17        & $+$0.27        & $+$0.25        & $+$0.22        & $-$1.19      & $+$0.11        \\
NGC~6540 & $-$0.28      & $+$1.03       & $+$0.34       & $+$0.19        & $+$0.14        & $+$0.24        & $+$0.26        & $-$1.06      & $+$0.05        \\
NGC~6642 & $-$0.14      & $+$0.78       & $+$0.45       & $+$0.28        & $+$0.16        & $+$0.32        & $+$0.25        & $-$1.13      & $+$0.17       \\
\hline

    \end{tabular}
\end{center}
\end{table*}

%
%                                                           Table
%%%%%%%%%%%%%%%%%%%%%%%%%%%%%%%%%%%%%%%%%%%%%%%%%%%%%%%%%%% END

\section{Discussion}
\label{section6}

Fig. \ref{figure3} plots [N/Fe] vs [C/Fe], [Al/Fe] vs [Si/Fe], [Ce/Fe] vs [Al/Fe], [Al/Fe] vs [Mg/Fe], [Si/Fe] vs [Mg/Fe], and [Ce/Fe] vs [N/Fe] diagrams for the sample stars, in order to check the expected anti-correlations for N-C and Al-Mg, correlations for Al-Si and Si-Mg \citep{Carretta2009, 2015Renzini, Meszaros2020}, and potential correlations for Ce-Al and Ce-N \citep[see, e.g.,][]{ngc6380, fernandez2022capos}.  We also show the re-calibrated abundances for 2 stars from \cite{barbuy2007} and 4 stars from \cite{Barbuy2018} (open squares and circles, respectively). Our star in common with \cite{ Barbuy2018} (filled circle) is depicted as a filled star symbol. The results appear to be consistent with each other, except for one star from \cite{Barbuy2018} in the Mg and Si abundances relative to Fe, which shows an evident shift in the Al-Si, Al-Mg, and Si-Mg diagrams. Despite the relatively few stars in our sample, anti-correlations for NGC~6558 might be present in the N-C, but the anti-correlations in Al-Mg, as well as the correlations in the Al-Si, Si-Mg, and Ce-N diagrams, are not clear. On the other hand, a weak anti-correlation between Ce and Al could be present, but the sample is too small to be definitive.

Fig. \ref{figure3} also displays stars of the GCs NGC~6522, from \citep{Fernandez-Trincado2019NGC6522}, as well as the CAPOS clusters Djorg~2, HP~1, NGC~6540, and NGC~6642. These clusters have similar metallicities to NGC~6558  ($-$1.2<$\langle$[Fe/H]$\rangle$<$-$1.0), as reported in \cite{Geisler2021}. Abundances derived from  \cite{Geisler2021} are from the \texttt{ASPCAP} system. To avoid any systematic errors and ensure consistency with the abundances measured with BACCHUS, we applied the same correction in [Fe/H] determined for NGC~6558 (as indicated by the dashed circles in the zoomed box of the top-right panel in Fig. \ref{figure1}).

We found that the NGC~6558 stars do not extend over a large range in [Al/Fe], but instead lie within a confined zone below [Al/Fe]=$+$0.3. This limit has been used as an easy and graphic marker between the first- and second-generation stars in GCs \citep{meszaros2015,Meszaros2020}. In spite of that, the high values in N abundance for the stars of NGC~6558 in Fig. \ref{figure3}, and the low values in C, are together a good indicator of multiple-population stars.  Thus, using only the Al abundance to distinguish between first- and second-generation stars may not be enough. Such enhanced N abundances exceed 4$\sigma$ above the typical Galactic level (dotted grey line at [N/Fe]$\sim +$0.81) at a metallicity of [Fe/H]=$-$1.15), as is shown in \cite{fernandez2022galactic}, and represented as a dashed line in Fig. \ref{fig:Polyplot}. This figure presents a 4$\sigma$ fourth-order polynomial fit above the mean behaviour of the nitrogen for the Galactic stars drawn from the APOGEE pipeline ASPCAP \citep{perez2016aspcap}, and corrected to BACCHUS abundances using the stars belonging to GCs in \cite[][grey dots]{Meszaros2020}. This polynomial fit was used by \cite{fernandez2022galactic} to select the nitrogen-enriched red giant stars distributed across the bulge, metal-poor disk, and halo of our Galaxy, which exhibit elemental
abundances comparable to those exclusively seen in GCs. These stars are thought to have been likely formed in the cluster and subsequently lost. 

All the BGCs displayed in Fig. \ref{figure3} exhibit a significant gap of approximately 0.3\,dex in nitrogen abundance, with the exception of NGC~6558 and NGC~6522. The latter two clusters, due to the small sample, only presented stars with [N/Fe] values over the Galactic levels within this range of metallicities. When compared to \citet[][their Fig. 17]{Meszaros2020}, this gap is solely observed in NGC~288, and barely indicated in NGC~5904. This suggests that the observed effect is likely due to the limited sample size in our clusters.

We could only measure the abundance of the $s$-process element Ce relative to Fe for three stars, as well as to N. Despite this, [Ce/Fe] shows values that are slightly overabundant compared to the Sun, but as expected by other BGCs at the same metallicity, as can be seen in the Ce-Al and Ce-N diagrams.

The remarkable similarities observed in the chemical behaviour between NGC~6558 and the other CAPOS BGCs with the same metallicity can be seen clearly in Fig. \ref{figure4} and Table \ref{tab:OtherClus}. Violin plots displaying the abundance distributions of NGC~6558 in comparison to NGC~6522, Djorg~2, HP~1, NGC~6540, and NGC~6642 are presented. All the BGC abundance distributions are strongly consistent with each other. The mean values of [N/Fe] for NGC~6558 and NGC~6522 are slightly greater than the others, primarily due to the small sample (5 stars for NGC~6522), which include only bona fide second-generation stars. On the other hand, NGC~6522 exhibits systematically higher Al abundances compared to other CAPOS clusters with the same metallicity. In fact, according to \citep{Fernandez-Trincado2019NGC6522}, this cluster exhibits a significant scatter in [Al/Fe], with a variation of $\sim$1\,dex when considering values from the literature.

Figs. \ref{figure3} and \ref{figure4} show that the [Mg/Fe] abundances for all these bulge clusters, including NGC~6558, do not exceed the value $\sim +$0.4. This observation is consistent with the values observed for the APOGEE bulge field stars \citep[see, for example,][]{Rojas-Arriagada2020,Geisler2021}. However, they differ from M~4, M~12, and NGC~288, whose average $\langle$[Mg/Fe]$\rangle$ values are around $\approx +$0.5 \citep{Meszaros2020}. These latter GCs share similar metallicities to NGC~6558, but they are not associated with the GB; instead, they are associated with disk formation (M~4 and M~12) or Gaia-Enceladus \citep[NGC~288,][]{Massari2019}.

%
%                                                           Section
%%%%%%%%%%%%%%%%%%%%%%%%%%%%%%%%%%%%%%%%%%%%%%%%%%%%%%%%%%% Conclusion
\section{Concluding remarks}
\label{section7}	

We have presented a new high-resolution (R$\sim$22,000) spectral analysis in the $H$-band for six members of the relatively old, moderately metal-poor BGC NGC~6558, obtained as part of the CAPOS survey.  The memberships are very likely, due to their spatial distribution, Gaia DR3 proper motions, and our own RV and metallicity from APOGEE spectra. We found a mean heliocentric RV of $V_r$ = $-$193.4 $\pm$ 0.9 km\,s$^{-1}$, where the error is the standard error of the mean, in good agreement with \cite{Barbuy2018}. Out of the six stars, four presented spectral S/N greater than 70 per pixel, which allowed us to determine their abundance ratios [C/Fe], [N/Fe], [O/Fe], [Mg/Fe], [Al/Fe], [Si/Fe], [Ca/Fe], [Ti/Fe], [Fe/H], [Ni/Fe], and [Ce/Fe] in a reliable way, through use of the \texttt{BACCHUS} code.

We found a mean metallicity of $\langle$[Fe/H]$\rangle$ = $-1.15$ for NGC~6558, in good agreement with \cite{Barbuy2018} when their value is scaled to the same Solar iron abundance as ours. We also found mean values of the $\alpha$-(O, Mg, Si, Ca, Ti) elements that are slightly enhanced, iron-peak (Ni) essentially Solar, and the $s$-process element (Ce) slightly overabundant compared to the Sun. The high enrichment in nitrogen, together with the depleted carbon, indicates the presence of the multiple-population phenomenon in this cluster, although the Al values are below the \cite{Meszaros2020} limit for second-generation stars. This is the first time that the phenomenon of multiple-populations is observed for this cluster. Such high nitrogen enrichment for NGC~6558 agrees with the nitrogen abundances for the CAPOS BGCs at the same metallicities, turning them into potential progenitors of the nitrogen-enhanced moderately metal-poor field stars identified in the inner bulge region \citep[e.g.,][]{trincado2016, trincado2017, schiavon2017StarsApogee, Trincado2019MNRAS, trincado2020dynamical, fernandez2021,fernandez2022capos}.

Despite our small statistical sample, we have observed consistent abundance patterns in all chemical elements studied in this work for all CAPOS BGCs in the same metallicity range as NGC~6558 ($-$1.2<$\langle$[Fe/H]$\rangle$<$-$1.0). However, NGC~6522 stands out due to its significant spread in [Al/Fe]. These clusters present nitrogen enhancement and carbon depletion, indicating the presence of multiple populations. Furthermore, the [Mg/Fe] abundances  in these clusters remain within the range observed for bulge field stars, in contrast to some Galactic GCs outside the bulge at the same metallicities.

\section*{Acknowledgements}

D.G-D and J.G.F-T gratefully acknowledge support from the Joint Committee ESO-Government of Chile 2021 (ORP 023/2021). J.G.F-T also acknowledges the financial support provided by Agencia Nacional de Investigaci\'on y Desarrollo de Chile (ANID) under the Proyecto Fondecyt Iniciaci\'on 2022 Agreement No. 11220340, and from ANID under the Concurso de Fomento a la Vinculaci\'on Internacional para Instituciones de Investigaci\'on Regionales (Modalidad corta duraci\'on) Agreement No. FOVI210020, and from Becas Santander Movilidad Internacional Profesores 2022, Banco Santander Chile, and from Vicerrector\'ia de Investigaci\'on y Desarrollo Tecnol\'ogico (VRIDT) at Universidad Cat\'olica del Norte (UCN) under resoluci\'on No 061/2022-VRIDT-UCN. D.G. gratefully acknowledges support from the ANID BASAL project ACE210002. D.G. also acknowledges financial support from the Direcci\'on de Investigaci\'on y Desarrollo de la Universidad de La Serena through the Programa de Incentivo a la Investigaci\'on de Acad\'emicos (PIA-DIDULS). S.V. gratefully acknowledges the support provided by Fondcyt regular n. 1220264. S.V. gratefully acknowledges support by the ANID BASAL projects ACE210002 and FB210003. D.M. gratefully acknowledges support by the ANID BASAL projects ACE210002 and FB210003 and by Fondecyt project No. 1220724. DM Also acknowledges support from CNPq/Brazil through project 350104/2022-0. T.C.B. acknowledges partial support for this work from grant PHY 14-30152; Physics Frontier Center/JINA Center for the Evolution of the Elements (JINA-CEE), and from OISE-1927130: The International Research Network for Nuclear Astrophysics (IReNA), awarded by the U.S. National Science Foundation. B.T. gratefully acknowledges support from the Natural Science Foundation of Guangdong Province under grant No. 2022A1515010732, and the National Natural Science Foundation of China through grants No. 12233013. BB acknowledges partial financial support from FAPESP, CNPq and Capes Financial code 001. C.M. thanks the support provided by FONDECYT Postdoctorado No.3210144.
A.M. gratefully acknowledges support by the ANID BASAL project FB210003,  FONDECYT Regular grant 1212046, and funding from the Max Planck Society through a “PartnerGroup” grant.

Funding for the Sloan Digital Sky Survey IV has been provided by the Alfred P. Sloan Foundation, the U.S. Department of Energy Office of Science, and the Participating Institutions. SDSS-IV acknowledges support and resources from the Center for High-Performance Computing at the University of Utah. The SDSS website is www.sdss.org. SDSS-IV is managed by the Astrophysical Research Consortium for the Participating Institutions of the SDSS Collaboration including the Brazilian Participation Group, the Carnegie Institution for Science, Carnegie Mellon University, the Chilean Participation Group, the French Participation Group, Harvard-Smithsonian Center for Astrophysics, Instituto de Astrof\`{i}sica de Canarias, The Johns Hopkins University, Kavli Institute for the Physics and Mathematics of the Universe (IPMU) / University of Tokyo, Lawrence Berkeley National Laboratory, Leibniz Institut f\"{u}r Astrophysik Potsdam (AIP), Max-Planck-Institut f\"{u}r Astronomie (MPIA Heidelberg), Max-Planck-Institut f\"{u}r Astrophysik (MPA Garching), Max-Planck-Institut f\"{u}r Extraterrestrische Physik (MPE), National Astronomical Observatory of China, New Mexico State University, New York University, University of Notre Dame, Observat\'{o}rio Nacional / MCTI, The Ohio State University, Pennsylvania State University, Shanghai Astronomical Observatory, United Kingdom Participation Group, Universidad Nacional Aut\'{o}noma de M\'{e}xico, University of Arizona, University of Colorado Boulder, University of Oxford, University of Portsmouth, University of Utah, University of Virginia, University of Washington, University of Wisconsin, Vanderbilt University, and Yale University.\\
This work has made use of data from the European Space Agency (ESA) mission \textit{Gaia} (\url{http://www.cosmos.esa.int/gaia}), processed by the \textit{Gaia} Data Processing and Analysis Consortium (DPAC, \url{http://www.cosmos.esa.int/web/gaia/dpac/consortium}). Funding for the DPAC has been provided by national institutions, in particular, the institutions participating in the \textit{Gaia} Multilateral Agreement.\\

%%%%%%%%%%%%%%%%%%%%%%%%%%%%%%%%%%%%%%%%%%%%%%%%%%
\section*{Data Availability}

All APOGEE DR~17 data upon which this study is based are publicly available and can be found at \url{https://www.sdss4.org/dr17}.

%%%%%%%%%%%%%%%%%%%% REFERENCES %%%%%%%%%%%%%%%%%%

% The best way to enter references is to use BibTeX:

\bibliographystyle{mnras}
\bibliography{ngc6558} % if your bibtex file is called example.bib

\begin{thebibliography}{}
\makeatletter
\relax
\def\mn@urlcharsother{\let\do\@makeother \do\$\do\&\do\#\do\^\do\_\do\%\do\~}
\def\mn@doi{\begingroup\mn@urlcharsother \@ifnextchar [ {\mn@doi@} {\mn@doi@[]}}
\def\mn@doi@[#1]#2{\def\@tempa{#1}\ifx\@tempa\@empty \href {http://dx.doi.org/#2} {doi:#2}\else \href {http://dx.doi.org/#2} {#1}\fi \endgroup}
\def\mn@eprint#1#2{\mn@eprint@#1:#2::\@nil}
\def\mn@eprint@arXiv#1{\href {http://arxiv.org/abs/#1} {{\tt arXiv:#1}}}
\def\mn@eprint@dblp#1{\href {http://dblp.uni-trier.de/rec/bibtex/#1.xml} {dblp:#1}}
\def\mn@eprint@#1:#2:#3:#4\@nil{\def\@tempa {#1}\def\@tempb {#2}\def\@tempc {#3}\ifx \@tempc \@empty \let \@tempc \@tempb \let \@tempb \@tempa \fi \ifx \@tempb \@empty \def\@tempb {arXiv}\fi \@ifundefined {mn@eprint@\@tempb}{\@tempb:\@tempc}{\expandafter \expandafter \csname mn@eprint@\@tempb\endcsname \expandafter{\@tempc}}}

\bibitem[\protect\citeauthoryear{{Abdurro'uf} et~al.,}{{Abdurro'uf} et~al.}{2022a}]{Accetta2022}
{Abdurro'uf} et~al., 2022a, \mn@doi [\apjs] {10.3847/1538-4365/ac4414}, \href {https://ui.adsabs.harvard.edu/abs/2022ApJS..259...35A} {259, 35}

\bibitem[\protect\citeauthoryear{{Abdurro'uf} et~al.,}{{Abdurro'uf} et~al.}{2022b}]{Abdurro2022}
{Abdurro'uf} et~al., 2022b, \mn@doi [\apjs] {10.3847/1538-4365/ac4414}, \href {https://ui.adsabs.harvard.edu/abs/2022ApJS..259...35A} {259, 35}

\bibitem[\protect\citeauthoryear{{Ahumada} et~al.,}{{Ahumada} et~al.}{2020}]{Ahumada2020}
{Ahumada} R.,  et~al., 2020, \mn@doi [\apjs] {10.3847/1538-4365/ab929e}, \href {https://ui.adsabs.harvard.edu/abs/2020ApJS..249....3A} {249, 3}

\bibitem[\protect\citeauthoryear{{Allende Prieto}, {Beers}, {Wilhelm}, {Newberg}, {Rockosi}, {Yanny}  \& {Lee}}{{Allende Prieto} et~al.}{2006}]{Allende2006}
{Allende Prieto} C.,  {Beers} T.~C.,  {Wilhelm} R.,  {Newberg} H.~J.,  {Rockosi} C.~M.,  {Yanny} B.,   {Lee} Y.~S.,  2006, \mn@doi [\apj] {10.1086/498131}, \href {https://ui.adsabs.harvard.edu/abs/2006ApJ...636..804A} {636, 804}

\bibitem[\protect\citeauthoryear{{Alvarez} \& {Plez}}{{Alvarez} \& {Plez}}{1998}]{Alvarez1998}
{Alvarez} R.,  {Plez} B.,  1998, \aap, \href {https://ui.adsabs.harvard.edu/abs/1998A&A...330.1109A} {330, 1109}

\bibitem[\protect\citeauthoryear{{Asplund}, {Grevesse}  \& {Sauval}}{{Asplund} et~al.}{2005}]{Asplund2005}
{Asplund} M.,  {Grevesse} N.,   {Sauval} A.~J.,  2005, in {Barnes} Thomas~G. I.,  {Bash} F.~N.,  eds,  \pasp Vol. 336, Cosmic Abundances as Records of Stellar Evolution and Nucleosynthesis. p.~25

\bibitem[\protect\citeauthoryear{{Barbuy}, {Zoccali}, {Ortolani}, {Minniti}, {Hill}, {Renzini}, {Bica}  \& {G{\'o}mez}}{{Barbuy} et~al.}{2007}]{barbuy2007}
{Barbuy} B.,  {Zoccali} M.,  {Ortolani} S.,  {Minniti} D.,  {Hill} V.,  {Renzini} A.,  {Bica} E.,   {G{\'o}mez} A.,  2007, \mn@doi [\aj] {10.1086/521556}, \href {https://ui.adsabs.harvard.edu/abs/2007AJ....134.1613B} {134, 1613}

\bibitem[\protect\citeauthoryear{{Barbuy}, {Zoccali}, {Ortolani}, {Hill}, {Minniti}, {Bica}, {Renzini}  \& {G{\'o}mez}}{{Barbuy} et~al.}{2009}]{Barbuy2009}
{Barbuy} B.,  {Zoccali} M.,  {Ortolani} S.,  {Hill} V.,  {Minniti} D.,  {Bica} E.,  {Renzini} A.,   {G{\'o}mez} A.,  2009, \mn@doi [\aap] {10.1051/0004-6361/200912748}, \href {https://ui.adsabs.harvard.edu/abs/2009A&A...507..405B} {507, 405}

\bibitem[\protect\citeauthoryear{{Barbuy} et~al.,}{{Barbuy} et~al.}{2014}]{Barbuy2014}
{Barbuy} B.,  et~al., 2014, \mn@doi [\aap] {10.1051/0004-6361/201424311}, \href {https://ui.adsabs.harvard.edu/abs/2014A&A...570A..76B} {570, A76}

\bibitem[\protect\citeauthoryear{{Barbuy}, {Chiappini}  \& {Gerhard}}{{Barbuy} et~al.}{2018a}]{barbuy2018chemodynamical}
{Barbuy} B.,  {Chiappini} C.,   {Gerhard} O.,  2018a, \mn@doi [\araa] {10.1146/annurev-astro-081817-051826}, \href {https://ui.adsabs.harvard.edu/abs/2018ARA&A..56..223B} {56, 223}

\bibitem[\protect\citeauthoryear{{Barbuy} et~al.,}{{Barbuy} et~al.}{2018b}]{Barbuy2018}
{Barbuy} B.,  et~al., 2018b, \mn@doi [\aap] {10.1051/0004-6361/201833953}, \href {https://ui.adsabs.harvard.edu/abs/2018A&A...619A.178B} {619, A178}

\bibitem[\protect\citeauthoryear{{Baumgardt} \& {Vasiliev}}{{Baumgardt} \& {Vasiliev}}{2021a}]{2021-Baumgardt-Vasiliev}
{Baumgardt} H.,  {Vasiliev} E.,  2021a, \mn@doi [\mnras] {10.1093/mnras/stab1474}, \href {https://ui.adsabs.harvard.edu/abs/2021MNRAS.505.5957B} {505, 5957}

\bibitem[\protect\citeauthoryear{{Baumgardt} \& {Vasiliev}}{{Baumgardt} \& {Vasiliev}}{2021b}]{Baumgardt-Vasiliev2021}
{Baumgardt} H.,  {Vasiliev} E.,  2021b, \mn@doi [\mnras] {10.1093/mnras/stab1474}, \href {https://ui.adsabs.harvard.edu/abs/2021MNRAS.505.5957B} {505, 5957}

\bibitem[\protect\citeauthoryear{{Beaton} et~al.,}{{Beaton} et~al.}{2021}]{Beaton2021}
{Beaton} R.~L.,  et~al., 2021, \mn@doi [\aj] {10.3847/1538-3881/ac260c}, \href {https://ui.adsabs.harvard.edu/abs/2021AJ....162..302B} {162, 302}

\bibitem[\protect\citeauthoryear{{Blanton} et~al.,}{{Blanton} et~al.}{2017}]{Blanton2017}
{Blanton} M.~R.,  et~al., 2017, \mn@doi [\aj] {10.3847/1538-3881/aa7567}, \href {https://ui.adsabs.harvard.edu/abs/2017AJ....154...28B} {154, 28}

\bibitem[\protect\citeauthoryear{{Bowen} \& {Vaughan}}{{Bowen} \& {Vaughan}}{1973}]{Bowen1973}
{Bowen} I.~S.,  {Vaughan} A.~H. J.,  1973, \mn@doi [\ao] {10.1364/AO.12.001430}, \href {https://ui.adsabs.harvard.edu/abs/1973ApOpt..12.1430B} {12, 1430}

\bibitem[\protect\citeauthoryear{{Bressan}, {Marigo}, {Girardi}, {Salasnich}, {Dal Cero}, {Rubele}  \& {Nanni}}{{Bressan} et~al.}{2012}]{Bressan2012}
{Bressan} A.,  {Marigo} P.,  {Girardi} L.,  {Salasnich} B.,  {Dal Cero} C.,  {Rubele} S.,   {Nanni} A.,  2012, \mn@doi [\mnras] {10.1111/j.1365-2966.2012.21948.x}, \href {http://adsabs.harvard.edu/abs/2012MNRAS.427..127B} {427, 127}

\bibitem[\protect\citeauthoryear{{Cardelli}, {Clayton}  \& {Mathis}}{{Cardelli} et~al.}{1989}]{Cardelli1989}
{Cardelli} J.~A.,  {Clayton} G.~C.,   {Mathis} J.~S.,  1989, \mn@doi [\apj] {10.1086/167900}, \href {https://ui.adsabs.harvard.edu/abs/1989ApJ...345..245C} {345, 245}

\bibitem[\protect\citeauthoryear{{Carretta}, {Bragaglia}, {Gratton}  \& {Lucatello}}{{Carretta} et~al.}{2009}]{Carretta2009}
{Carretta} E.,  {Bragaglia} A.,  {Gratton} R.,   {Lucatello} S.,  2009, \mn@doi [\aap] {10.1051/0004-6361/200912097}, \href {https://ui.adsabs.harvard.edu/abs/2009A&A...505..139C} {505, 139}

\bibitem[\protect\citeauthoryear{{Cescutti}}{{Cescutti}}{2008}]{Cescutti2008}
{Cescutti} G.,  2008, \mn@doi [\aap] {10.1051/0004-6361:20078571}, \href {https://ui.adsabs.harvard.edu/abs/2008A&A...481..691C} {481, 691}

\bibitem[\protect\citeauthoryear{{Chiappini}, {Frischknecht}, {Meynet}, {Hirschi}, {Barbuy}, {Pignatari}, {Decressin}  \& {Maeder}}{{Chiappini} et~al.}{2011}]{chiappini2011}
{Chiappini} C.,  {Frischknecht} U.,  {Meynet} G.,  {Hirschi} R.,  {Barbuy} B.,  {Pignatari} M.,  {Decressin} T.,   {Maeder} A.,  2011, \mn@doi [\nat] {10.1038/nature10000}, \href {https://ui.adsabs.harvard.edu/abs/2011Natur.472..454C} {472, 454}

\bibitem[\protect\citeauthoryear{{Cohen}, {Bellini}, {Casagrande}, {Brown}, {Correnti}  \& {Kalirai}}{{Cohen} et~al.}{2021}]{cohen2021}
{Cohen} R.~E.,  {Bellini} A.,  {Casagrande} L.,  {Brown} T.~M.,  {Correnti} M.,   {Kalirai} J.~S.,  2021, \mn@doi [\aj] {10.3847/1538-3881/ac281f}, \href {https://ui.adsabs.harvard.edu/abs/2021AJ....162..228C} {162, 228}

\bibitem[\protect\citeauthoryear{{Cunha} et~al.,}{{Cunha} et~al.}{2017}]{Cunha2017}
{Cunha} K.,  et~al., 2017, \mn@doi [\apj] {10.3847/1538-4357/aa7beb}, \href {https://ui.adsabs.harvard.edu/abs/2017ApJ...844..145C} {844, 145}

\bibitem[\protect\citeauthoryear{{Ernandes}, {Barbuy}, {Alves-Brito}, {Fria{\c{c}}a}, {Siqueira-Mello}  \& {Allen}}{{Ernandes} et~al.}{2018}]{ernandes2018}
{Ernandes} H.,  {Barbuy} B.,  {Alves-Brito} A.,  {Fria{\c{c}}a} A.,  {Siqueira-Mello} C.,   {Allen} D.~M.,  2018, \mn@doi [\aap] {10.1051/0004-6361/201731708}, \href {https://ui.adsabs.harvard.edu/abs/2018A&A...616A..18E} {616, A18}

\bibitem[\protect\citeauthoryear{{Fern{\'a}ndez-Trincado} et~al.,}{{Fern{\'a}ndez-Trincado} et~al.}{2016}]{trincado2016}
{Fern{\'a}ndez-Trincado} J.~G.,  et~al., 2016, \mn@doi [\apj] {10.3847/1538-4357/833/2/132}, \href {https://ui.adsabs.harvard.edu/abs/2016ApJ...833..132F} {833, 132}

\bibitem[\protect\citeauthoryear{{Fern{\'a}ndez-Trincado} et~al.,}{{Fern{\'a}ndez-Trincado} et~al.}{2017}]{trincado2017}
{Fern{\'a}ndez-Trincado} J.~G.,  et~al., 2017, \mn@doi [\apjl] {10.3847/2041-8213/aa8032}, \href {https://ui.adsabs.harvard.edu/abs/2017ApJ...846L...2F} {846, L2}

\bibitem[\protect\citeauthoryear{{Fern{\'a}ndez-Trincado}, {Beers}, {Tang}, {Moreno}, {P{\'e}rez-Villegas}  \& {Ortigoza-Urdaneta}}{{Fern{\'a}ndez-Trincado} et~al.}{2019a}]{Trincado2019MNRAS}
{Fern{\'a}ndez-Trincado} J.~G.,  {Beers} T.~C.,  {Tang} B.,  {Moreno} E.,  {P{\'e}rez-Villegas} A.,   {Ortigoza-Urdaneta} M.,  2019a, \mn@doi [\mnras] {10.1093/mnras/stz1848}, \href {https://ui.adsabs.harvard.edu/abs/2019MNRAS.488.2864F} {488, 2864}

\bibitem[\protect\citeauthoryear{{Fern{\'a}ndez-Trincado} et~al.,}{{Fern{\'a}ndez-Trincado} et~al.}{2019b}]{Trincado2019}
{Fern{\'a}ndez-Trincado} J.~G.,  et~al., 2019b, \mn@doi [\aap] {10.1051/0004-6361/201834391}, \href {https://ui.adsabs.harvard.edu/abs/2019A&A...627A.178F} {627, A178}

\bibitem[\protect\citeauthoryear{{Fern{\'a}ndez-Trincado} et~al.,}{{Fern{\'a}ndez-Trincado} et~al.}{2019c}]{Fernandez-Trincado2019NGC6522}
{Fern{\'a}ndez-Trincado} J.~G.,  et~al., 2019c, \mn@doi [\aap] {10.1051/0004-6361/201834391}, \href {https://ui.adsabs.harvard.edu/abs/2019A&A...627A.178F} {627, A178}

\bibitem[\protect\citeauthoryear{{Fern{\'a}ndez-Trincado}, {Chaves-Velasquez}, {P{\'e}rez-Villegas}, {Vieira}, {Moreno}, {Ortigoza-Urdaneta}  \& {Vega-Neme}}{{Fern{\'a}ndez-Trincado} et~al.}{2020a}]{trincado2020dynamical}
{Fern{\'a}ndez-Trincado} J.~G.,  {Chaves-Velasquez} L.,  {P{\'e}rez-Villegas} A.,  {Vieira} K.,  {Moreno} E.,  {Ortigoza-Urdaneta} M.,   {Vega-Neme} L.,  2020a, \mn@doi [\mnras] {10.1093/mnras/staa1386}, \href {https://ui.adsabs.harvard.edu/abs/2020MNRAS.495.4113F} {495, 4113}

\bibitem[\protect\citeauthoryear{{Fern{\'a}ndez-Trincado}, {Beers}, {Minniti}, {Tang}, {Villanova}, {Geisler}, {P{\'e}rez-Villegas}  \& {Vieira}}{{Fern{\'a}ndez-Trincado} et~al.}{2020b}]{BacchuAspcap2020}
{Fern{\'a}ndez-Trincado} J.~G.,  {Beers} T.~C.,  {Minniti} D.,  {Tang} B.,  {Villanova} S.,  {Geisler} D.,  {P{\'e}rez-Villegas} A.,   {Vieira} K.,  2020b, \mn@doi [\aap] {10.1051/0004-6361/202039207}, \href {https://ui.adsabs.harvard.edu/abs/2020A&A...643L...4F} {643, L4}

\bibitem[\protect\citeauthoryear{{Fern{\'a}ndez-Trincado} et~al.,}{{Fern{\'a}ndez-Trincado} et~al.}{2020c}]{trincadoUKS1}
{Fern{\'a}ndez-Trincado} J.~G.,  et~al., 2020c, \mn@doi [\aap] {10.1051/0004-6361/202039328}, \href {https://ui.adsabs.harvard.edu/abs/2020A&A...643A.145F} {643, A145}

\bibitem[\protect\citeauthoryear{{Fern{\'a}ndez-Trincado} et~al.,}{{Fern{\'a}ndez-Trincado} et~al.}{2021a}]{fernandez2021}
{Fern{\'a}ndez-Trincado} J.~G.,  et~al., 2021a, \mn@doi [\aap] {10.1051/0004-6361/202040255}, \href {https://ui.adsabs.harvard.edu/abs/2021A&A...647A..64F} {647, A64}

\bibitem[\protect\citeauthoryear{{Fern{\'a}ndez-Trincado} et~al.,}{{Fern{\'a}ndez-Trincado} et~al.}{2021b}]{trincadoSgr}
{Fern{\'a}ndez-Trincado} J.~G.,  et~al., 2021b, \mn@doi [\aap] {10.1051/0004-6361/202140306}, \href {https://ui.adsabs.harvard.edu/abs/2021A&A...648A..70F} {648, A70}

\bibitem[\protect\citeauthoryear{{Fern{\'a}ndez-Trincado} et~al.,}{{Fern{\'a}ndez-Trincado} et~al.}{2021c}]{trincadoVVVCL001}
{Fern{\'a}ndez-Trincado} J.~G.,  et~al., 2021c, \mn@doi [\apjl] {10.3847/2041-8213/abdf47}, \href {https://ui.adsabs.harvard.edu/abs/2021ApJ...908L..42F} {908, L42}

\bibitem[\protect\citeauthoryear{{Fern{\'a}ndez-Trincado} et~al.,}{{Fern{\'a}ndez-Trincado} et~al.}{2021d}]{ngc6380}
{Fern{\'a}ndez-Trincado} J.~G.,  et~al., 2021d, \mn@doi [\apjl] {10.3847/2041-8213/ac1c7e}, \href {https://ui.adsabs.harvard.edu/abs/2021ApJ...918L...9F} {918, L9}

\bibitem[\protect\citeauthoryear{Fern\'andez-Trincado, {Minniti, Dante}, {Garro, Elisa R.}  \& {Villanova, Sandro}}{Fern\'andez-Trincado et~al.}{2022a}]{Fernandez-Trincado2022}
Fern\'andez-Trincado J.~G.,  {Minniti, Dante} {Garro, Elisa R.}  {Villanova, Sandro} 2022a, \mn@doi [A\&A] {10.1051/0004-6361/202142222}, 657, A84

\bibitem[\protect\citeauthoryear{{Fern{\'a}ndez-Trincado} et~al.,}{{Fern{\'a}ndez-Trincado} et~al.}{2022b}]{fernandez2022capos}
{Fern{\'a}ndez-Trincado} J.~G.,  et~al., 2022b, \mn@doi [\aap] {10.1051/0004-6361/202141742}, \href {https://ui.adsabs.harvard.edu/abs/2022A&A...658A.116F} {658, A116}

\bibitem[\protect\citeauthoryear{{Fern{\'a}ndez-Trincado} et~al.,}{{Fern{\'a}ndez-Trincado} et~al.}{2022c}]{fernandez2022galactic}
{Fern{\'a}ndez-Trincado} J.~G.,  et~al., 2022c, \mn@doi [\aap] {10.1051/0004-6361/202243195}, \href {https://ui.adsabs.harvard.edu/abs/2022A&A...663A.126F} {663, A126}

\bibitem[\protect\citeauthoryear{{Frischknecht} et~al.,}{{Frischknecht} et~al.}{2016}]{frischknecht2016}
{Frischknecht} U.,  et~al., 2016, \mn@doi [\mnras] {10.1093/mnras/stv2723}, \href {https://ui.adsabs.harvard.edu/abs/2016MNRAS.456.1803F} {456, 1803}

\bibitem[\protect\citeauthoryear{{Gaia Collaboration} et~al.,}{{Gaia Collaboration} et~al.}{2018}]{GaiaDR2}
{Gaia Collaboration} et~al., 2018, \mn@doi [\aap] {10.1051/0004-6361/201833051}, \href {https://ui.adsabs.harvard.edu/abs/2018A&A...616A...1G} {616, A1}

\bibitem[\protect\citeauthoryear{{Gaia Collaboration} et~al.,}{{Gaia Collaboration} et~al.}{2022}]{GaiaDR3}
{Gaia Collaboration} et~al., 2022, arXiv e-prints, \href {https://ui.adsabs.harvard.edu/abs/2022arXiv220800211G} {p. arXiv:2208.00211}

\bibitem[\protect\citeauthoryear{{Garc{\'\i}a P{\'e}rez} et~al.,}{{Garc{\'\i}a P{\'e}rez} et~al.}{2016}]{Garcia2016}
{Garc{\'\i}a P{\'e}rez} A.~E.,  et~al., 2016, \mn@doi [\aj] {10.3847/0004-6256/151/6/144}, \href {https://ui.adsabs.harvard.edu/abs/2016AJ....151..144G} {151, 144}

\bibitem[\protect\citeauthoryear{{Geisler} et~al.,}{{Geisler} et~al.}{2021}]{Geisler2021}
{Geisler} D.,  et~al., 2021, \mn@doi [\aap] {10.1051/0004-6361/202140436}, \href {https://ui.adsabs.harvard.edu/abs/2021A&A...652A.157G} {652, A157}

\bibitem[\protect\citeauthoryear{{Goriely}, {Bauswein}  \& {Janka}}{{Goriely} et~al.}{2011}]{goriely2011}
{Goriely} S.,  {Bauswein} A.,   {Janka} H.-T.,  2011, \mn@doi [\apjl] {10.1088/2041-8205/738/2/L32}, \href {https://ui.adsabs.harvard.edu/abs/2011ApJ...738L..32G} {738, L32}

\bibitem[\protect\citeauthoryear{{Gratton}, {Carretta}  \& {Castelli}}{{Gratton} et~al.}{1996}]{gratton1996}
{Gratton} R.~G.,  {Carretta} E.,   {Castelli} F.,  1996, \mn@doi [\aap] {10.48550/arXiv.astro-ph/9603011}, \href {https://ui.adsabs.harvard.edu/abs/1996A&A...314..191G} {314, 191}

\bibitem[\protect\citeauthoryear{{Grevesse} \& {Sauval}}{{Grevesse} \& {Sauval}}{1998}]{grevesse1998}
{Grevesse} N.,  {Sauval} A.~J.,  1998, \mn@doi [\ssr] {10.1023/A:1005161325181}, \href {https://ui.adsabs.harvard.edu/abs/1998SSRv...85..161G} {85, 161}

\bibitem[\protect\citeauthoryear{{Gunn} et~al.,}{{Gunn} et~al.}{2006}]{Gunn2006}
{Gunn} J.~E.,  et~al., 2006, \mn@doi [\aj] {10.1086/500975}, \href {https://ui.adsabs.harvard.edu/abs/2006AJ....131.2332G} {131, 2332}

\bibitem[\protect\citeauthoryear{{Gustafsson}, {Edvardsson}, {Eriksson}, {J{\o}rgensen}, {Nordlund}  \& {Plez}}{{Gustafsson} et~al.}{2008}]{Gustafsson2008}
{Gustafsson} B.,  {Edvardsson} B.,  {Eriksson} K.,  {J{\o}rgensen} U.~G.,  {Nordlund} {\AA}.,   {Plez} B.,  2008, \mn@doi [\aap] {10.1051/0004-6361:200809724}, \href {http://adsabs.harvard.edu/abs/2008A%26A...486..951G} {486, 951}

\bibitem[\protect\citeauthoryear{{Hansen} et~al.,}{{Hansen} et~al.}{2017}]{hansen2017}
{Hansen} T.~T.,  et~al., 2017, \mn@doi [\apj] {10.3847/1538-4357/aa634a}, \href {https://ui.adsabs.harvard.edu/abs/2017ApJ...838...44H} {838, 44}

\bibitem[\protect\citeauthoryear{{Harris}}{{Harris}}{1996}]{Harris1996}
{Harris} W.~E.,  1996, \mn@doi [\aj] {10.1086/118116}, \href {https://ui.adsabs.harvard.edu/abs/1996AJ....112.1487H} {112, 1487}

\bibitem[\protect\citeauthoryear{{Harris}}{{Harris}}{2010}]{Harris2010}
{Harris} W.~E.,  2010, arXiv e-prints, \href {https://ui.adsabs.harvard.edu/abs/2010arXiv1012.3224H} {p. arXiv:1012.3224}

\bibitem[\protect\citeauthoryear{{Hasselquist} et~al.,}{{Hasselquist} et~al.}{2016}]{Hasselquist2016}
{Hasselquist} S.,  et~al., 2016, \mn@doi [\apj] {10.3847/1538-4357/833/1/81}, \href {https://ui.adsabs.harvard.edu/abs/2016ApJ...833...81H} {833, 81}

\bibitem[\protect\citeauthoryear{{Hawkins}, {Masseron}, {Jofr{\'e}}, {Gilmore}, {Elsworth}  \& {Hekker}}{{Hawkins} et~al.}{2016}]{Hawkins16}
{Hawkins} K.,  {Masseron} T.,  {Jofr{\'e}} P.,  {Gilmore} G.,  {Elsworth} Y.,   {Hekker} S.,  2016, \mn@doi [\aap] {10.1051/0004-6361/201628812}, \href {https://ui.adsabs.harvard.edu/abs/2016A&A...594A..43H} {594, A43}

\bibitem[\protect\citeauthoryear{{Heger}, {Fryer}, {Woosley}, {Langer}  \& {Hartmann}}{{Heger} et~al.}{2003}]{heger2003}
{Heger} A.,  {Fryer} C.~L.,  {Woosley} S.~E.,  {Langer} N.,   {Hartmann} D.~H.,  2003, \mn@doi [\apj] {10.1086/375341}, \href {https://ui.adsabs.harvard.edu/abs/2003ApJ...591..288H} {591, 288}

\bibitem[\protect\citeauthoryear{{Holtzman} et~al.,}{{Holtzman} et~al.}{2018}]{Holtzman2018}
{Holtzman} J.~A.,  et~al., 2018, \mn@doi [\aj] {10.3847/1538-3881/aad4f9}, \href {https://ui.adsabs.harvard.edu/abs/2018AJ....156..125H} {156, 125}

\bibitem[\protect\citeauthoryear{{Janka}}{{Janka}}{2012}]{janka2012}
{Janka} H.-T.,  2012, \mn@doi [Annual Review of Nuclear and Particle Science] {10.1146/annurev-nucl-102711-094901}, \href {https://ui.adsabs.harvard.edu/abs/2012ARNPS..62..407J} {62, 407}

\bibitem[\protect\citeauthoryear{{Ji}, {Frebel}, {Simon}  \& {Chiti}}{{Ji} et~al.}{2016}]{ji2016}
{Ji} A.~P.,  {Frebel} A.,  {Simon} J.~D.,   {Chiti} A.,  2016, \mn@doi [\apj] {10.3847/0004-637X/830/2/93}, \href {https://ui.adsabs.harvard.edu/abs/2016ApJ...830...93J} {830, 93}

\bibitem[\protect\citeauthoryear{{J{\"o}nsson} et~al.,}{{J{\"o}nsson} et~al.}{2018}]{Henrik2018}
{J{\"o}nsson} H.,  et~al., 2018, \mn@doi [\aj] {10.3847/1538-3881/aad4f5}, \href {https://ui.adsabs.harvard.edu/abs/2018AJ....156..126J} {156, 126}

\bibitem[\protect\citeauthoryear{{J{\"o}nsson} et~al.,}{{J{\"o}nsson} et~al.}{2020}]{Henrik2020}
{J{\"o}nsson} H.,  et~al., 2020, \mn@doi [\aj] {10.3847/1538-3881/aba592}, \href {https://ui.adsabs.harvard.edu/abs/2020AJ....160..120J} {160, 120}

\bibitem[\protect\citeauthoryear{{Majewski} et~al.,}{{Majewski} et~al.}{2017}]{Majewski2017}
{Majewski} S.~R.,  et~al., 2017, \mn@doi [\aj] {10.3847/1538-3881/aa784d}, \href {https://ui.adsabs.harvard.edu/abs/2017AJ....154...94M} {154, 94}

\bibitem[\protect\citeauthoryear{{Malhan} et~al.,}{{Malhan} et~al.}{2022}]{malhan2022}
{Malhan} K.,  et~al., 2022, \mn@doi [\apj] {10.3847/1538-4357/ac4d2a}, \href {https://ui.adsabs.harvard.edu/abs/2022ApJ...926..107M} {926, 107}

\bibitem[\protect\citeauthoryear{{Massari}, {Koppelman}  \& {Helmi}}{{Massari} et~al.}{2019}]{Massari2019}
{Massari} D.,  {Koppelman} H.~H.,   {Helmi} A.,  2019, \mn@doi [\aap] {10.1051/0004-6361/201936135}, \href {https://ui.adsabs.harvard.edu/abs/2019A&A...630L...4M} {630, L4}

\bibitem[\protect\citeauthoryear{{Masseron}, {Merle}  \& {Hawkins}}{{Masseron} et~al.}{2016}]{Masseron2016}
{Masseron} T.,  {Merle} T.,   {Hawkins} K.,  2016, {BACCHUS: Brussels Automatic Code for Characterizing High accUracy Spectra} (\mn@eprint {ascl} {1605.004})

\bibitem[\protect\citeauthoryear{{M{\'e}sz{\'a}ros} et~al.,}{{M{\'e}sz{\'a}ros} et~al.}{2015}]{meszaros2015}
{M{\'e}sz{\'a}ros} S.,  et~al., 2015, \mn@doi [\aj] {10.1088/0004-6256/149/5/153}, \href {https://ui.adsabs.harvard.edu/abs/2015AJ....149..153M} {149, 153}

\bibitem[\protect\citeauthoryear{{M{\'e}sz{\'a}ros} et~al.,}{{M{\'e}sz{\'a}ros} et~al.}{2020}]{Meszaros2020}
{M{\'e}sz{\'a}ros} S.,  et~al., 2020, \mn@doi [\mnras] {10.1093/mnras/stz3496}, \href {https://ui.adsabs.harvard.edu/abs/2020MNRAS.492.1641M} {492, 1641}

\bibitem[\protect\citeauthoryear{{Minniti}}{{Minniti}}{1995}]{minniti1995}
{Minniti} D.,  1995, \mn@doi [\aj] {10.1086/117393}, \href {https://ui.adsabs.harvard.edu/abs/1995AJ....109.1663M} {109, 1663}

\bibitem[\protect\citeauthoryear{{Minniti} et~al.,}{{Minniti} et~al.}{2010}]{minniti2010}
{Minniti} D.,  et~al., 2010, \mn@doi [\na] {10.1016/j.newast.2009.12.002}, \href {https://ui.adsabs.harvard.edu/abs/2010NewA...15..433M} {15, 433}

\bibitem[\protect\citeauthoryear{{Moreno}, {Fern{\'a}ndez-Trincado}, {P{\'e}rez-Villegas}, {Chaves-Velasquez}  \& {Schuster}}{{Moreno} et~al.}{2022}]{moreno2022}
{Moreno} E.,  {Fern{\'a}ndez-Trincado} J.~G.,  {P{\'e}rez-Villegas} A.,  {Chaves-Velasquez} L.,   {Schuster} W.~J.,  2022, \mn@doi [\mnras] {10.1093/mnras/stab3724}, \href {https://ui.adsabs.harvard.edu/abs/2022MNRAS.510.5945M} {510, 5945}

\bibitem[\protect\citeauthoryear{{Nidever} et~al.,}{{Nidever} et~al.}{2015}]{Nidever2015}
{Nidever} D.~L.,  et~al., 2015, \mn@doi [\aj] {10.1088/0004-6256/150/6/173}, \href {https://ui.adsabs.harvard.edu/abs/2015AJ....150..173N} {150, 173}

\bibitem[\protect\citeauthoryear{{O'Donnell}}{{O'Donnell}}{1994}]{Donnell1994}
{O'Donnell} J.~E.,  1994, \mn@doi [\apj] {10.1086/173713}, \href {https://ui.adsabs.harvard.edu/abs/1994ApJ...422..158O} {422, 158}

\bibitem[\protect\citeauthoryear{{P{\'e}rez-Villegas}, {Barbuy}, {Kerber}, {Ortolani}, {Souza}  \& {Bica}}{{P{\'e}rez-Villegas} et~al.}{2020}]{Perez-Villegas2020}
{P{\'e}rez-Villegas} A.,  {Barbuy} B.,  {Kerber} L.~O.,  {Ortolani} S.,  {Souza} S.~O.,   {Bica} E.,  2020, \mn@doi [\mnras] {10.1093/mnras/stz3162}, \href {https://ui.adsabs.harvard.edu/abs/2020MNRAS.491.3251P} {491, 3251}

\bibitem[\protect\citeauthoryear{P{\'e}rez et~al.,}{P{\'e}rez et~al.}{2016}]{perez2016aspcap}
P{\'e}rez A. E.~G.,  et~al., 2016, \mn@doi [\aj] {10.3847/0004-6256/151/6/144}, \href {https://ui.adsabs.harvard.edu/abs/2016AJ....151..144G} {151, 144}

\bibitem[\protect\citeauthoryear{{Plez}}{{Plez}}{2012}]{Plez2012}
{Plez} B.,  2012, {Turbospectrum: Code for spectral synthesis}, Astrophysics Source Code Library, record ascl:1205.004 (\mn@eprint {ascl} {1205.004})

\bibitem[\protect\citeauthoryear{{Reichert}, {Hansen}, {Hanke}, {Sk{\'u}lad{\'o}ttir}, {Arcones}  \& {Grebel}}{{Reichert} et~al.}{2020}]{reichert2020}
{Reichert} M.,  {Hansen} C.~J.,  {Hanke} M.,  {Sk{\'u}lad{\'o}ttir} {\'A}.,  {Arcones} A.,   {Grebel} E.~K.,  2020, \mn@doi [\aap] {10.1051/0004-6361/201936930}, \href {https://ui.adsabs.harvard.edu/abs/2020A&A...641A.127R} {641, A127}

\bibitem[\protect\citeauthoryear{{Renzini} et~al.,}{{Renzini} et~al.}{2015}]{2015Renzini}
{Renzini} A.,  et~al., 2015, \mn@doi [\mnras] {10.1093/mnras/stv2268}, \href {https://ui.adsabs.harvard.edu/abs/2015MNRAS.454.4197R} {454, 4197}

\bibitem[\protect\citeauthoryear{{Rich}, {Ortolani}, {Bica}  \& {Barbuy}}{{Rich} et~al.}{1998}]{rich1998}
{Rich} R.~M.,  {Ortolani} S.,  {Bica} E.,   {Barbuy} B.,  1998, \mn@doi [\aj] {10.1086/300497}, \href {https://ui.adsabs.harvard.edu/abs/1998AJ....116.1295R} {116, 1295}

\bibitem[\protect\citeauthoryear{{Roederer} et~al.,}{{Roederer} et~al.}{2016}]{roederer2016}
{Roederer} I.~U.,  et~al., 2016, \mn@doi [\aj] {10.3847/0004-6256/151/3/82}, \href {https://ui.adsabs.harvard.edu/abs/2016AJ....151...82R} {151, 82}

\bibitem[\protect\citeauthoryear{{Rojas-Arriagada} et~al.,}{{Rojas-Arriagada} et~al.}{2020}]{Rojas-Arriagada2020}
{Rojas-Arriagada} A.,  et~al., 2020, \mn@doi [\mnras] {10.1093/mnras/staa2807}, \href {https://ui.adsabs.harvard.edu/abs/2020MNRAS.499.1037R} {499, 1037}

\bibitem[\protect\citeauthoryear{{Rossi}, {Ortolani}, {Barbuy}, {Bica}  \& {Bonfanti}}{{Rossi} et~al.}{2015}]{rossi2015}
{Rossi} L.~J.,  {Ortolani} S.,  {Barbuy} B.,  {Bica} E.,   {Bonfanti} A.,  2015, \mn@doi [\mnras] {10.1093/mnras/stv748}, \href {https://ui.adsabs.harvard.edu/abs/2015MNRAS.450.3270R} {450, 3270}

\bibitem[\protect\citeauthoryear{{Saito} et~al.,}{{Saito} et~al.}{2012}]{Saito2012}
{Saito} R.~K.,  et~al., 2012, \mn@doi [\aap] {10.1051/0004-6361/201118407}, \href {https://ui.adsabs.harvard.edu/abs/2012A&A...537A.107S} {537, A107}

\bibitem[\protect\citeauthoryear{{Santana} et~al.,}{{Santana} et~al.}{2021}]{Santana2021}
{Santana} F.~A.,  et~al., 2021, \mn@doi [\aj] {10.3847/1538-3881/ac2cbc}, \href {https://ui.adsabs.harvard.edu/abs/2021AJ....162..303S} {162, 303}

\bibitem[\protect\citeauthoryear{{Savino}, {Koch}, {Prudil}, {Kunder}  \& {Smolec}}{{Savino} et~al.}{2020}]{Savino2020}
{Savino} A.,  {Koch} A.,  {Prudil} Z.,  {Kunder} A.,   {Smolec} R.,  2020, \mn@doi [\aap] {10.1051/0004-6361/202038305}, \href {https://ui.adsabs.harvard.edu/abs/2020A&A...641A..96S} {641, A96}

\bibitem[\protect\citeauthoryear{{Schiavon} et~al.,}{{Schiavon} et~al.}{2017}]{schiavon2017StarsApogee}
{Schiavon} R.~P.,  et~al., 2017, \mn@doi [\mnras] {10.1093/mnras/stw2162}, \href {https://ui.adsabs.harvard.edu/abs/2017MNRAS.465..501S} {465, 501}

\bibitem[\protect\citeauthoryear{{Shetrone} et~al.,}{{Shetrone} et~al.}{2015}]{Shetrone2015}
{Shetrone} M.,  et~al., 2015, \mn@doi [\apjs] {10.1088/0067-0049/221/2/24}, \href {https://ui.adsabs.harvard.edu/abs/2015ApJS..221...24S} {221, 24}

\bibitem[\protect\citeauthoryear{{Simon}}{{Simon}}{2019}]{simon2019}
{Simon} J.~D.,  2019, \mn@doi [\araa] {10.1146/annurev-astro-091918-104453}, \href {https://ui.adsabs.harvard.edu/abs/2019ARA&A..57..375S} {57, 375}

\bibitem[\protect\citeauthoryear{Skrutskie et~al.,}{Skrutskie et~al.}{2006}]{skrutskie2006}
Skrutskie M.,  et~al., 2006, The Astronomical Journal, 131, 1163

\bibitem[\protect\citeauthoryear{{Smith} et~al.,}{{Smith} et~al.}{2021}]{Smith2021}
{Smith} V.~V.,  et~al., 2021, arXiv e-prints, \href {https://ui.adsabs.harvard.edu/abs/2021arXiv210310112S} {p. arXiv:2103.10112}

\bibitem[\protect\citeauthoryear{{Trager}, {King}  \& {Djorgovski}}{{Trager} et~al.}{1995}]{Trager1995}
{Trager} S.~C.,  {King} I.~R.,   {Djorgovski} S.,  1995, \mn@doi [The Astronomical Journal] {10.1086/117268}, \href {https://ui.adsabs.harvard.edu/abs/1995AJ....109..218T} {109, 218}

\bibitem[\protect\citeauthoryear{{Wanajo}, {Hirai}  \& {Prantzos}}{{Wanajo} et~al.}{2021}]{wanajo2021}
{Wanajo} S.,  {Hirai} Y.,   {Prantzos} N.,  2021, \mn@doi [\mnras] {10.1093/mnras/stab1655}, \href {https://ui.adsabs.harvard.edu/abs/2021MNRAS.505.5862W} {505, 5862}

\bibitem[\protect\citeauthoryear{{Wilson} et~al.,}{{Wilson} et~al.}{2019}]{Wilson2019}
{Wilson} J.~C.,  et~al., 2019, \mn@doi [\pasp] {10.1088/1538-3873/ab0075}, \href {https://ui.adsabs.harvard.edu/abs/2019PASP..131e5001W} {131, 055001}

\bibitem[\protect\citeauthoryear{{Zamora} et~al.,}{{Zamora} et~al.}{2015}]{Zamora2015}
{Zamora} O.,  et~al., 2015, \mn@doi [\aj] {10.1088/0004-6256/149/6/181}, \href {https://ui.adsabs.harvard.edu/abs/2015AJ....149..181Z} {149, 181}

\bibitem[\protect\citeauthoryear{{Zasowski} et~al.,}{{Zasowski} et~al.}{2013}]{Zasowski2013}
{Zasowski} G.,  et~al., 2013, \mn@doi [\aj] {10.1088/0004-6256/146/4/81}, \href {https://ui.adsabs.harvard.edu/abs/2013AJ....146...81Z} {146, 81}

\bibitem[\protect\citeauthoryear{{Zasowski} et~al.,}{{Zasowski} et~al.}{2017}]{Zasowski2017}
{Zasowski} G.,  et~al., 2017, \mn@doi [\aj] {10.3847/1538-3881/aa8df9}, \href {https://ui.adsabs.harvard.edu/abs/2017AJ....154..198Z} {154, 198}

\makeatother
\end{thebibliography}

% Alternatively you could enter them by hand, like this:
% This method is tedious and prone to error if you have lots of references
%\begin{thebibliography}{99}
%\bibitem[\protect\citeauthoryear{Author}{2012}]{Author2012}
%Author A.~N., 2013, Journal of Improbable Astronomy, 1, 1
%\bibitem[\protect\citeauthoryear{Others}{2013}]{Others2013}
%Others S., 2012, Journal of Interesting Stuff, 17, 198
%\end{thebibliography}

%%%%%%%%%%%%%%%%%%%%%%%%%%%%%%%%%%%%%%%%%%%%%%%%%%

% Don't change these lines
\bsp	% typesetting comment
\label{lastpage}
\end{document}